\documentclass[10pt,letterpaper]{article}
\usepackage[top=0.85in,left=2.75in,footskip=0.75in]{geometry}

\usepackage{color}
\usepackage{ulem}

\usepackage{changepage}

\usepackage{textcomp,marvosym}

\usepackage{fixltx2e}

\usepackage{amsmath,amssymb}

\usepackage{cite}

\usepackage{nameref,hyperref}

\usepackage{microtype}
\DisableLigatures[f]{encoding = *, family = * }

\usepackage{siunitx}

\raggedright
\usepackage[aboveskip=1pt,labelfont=bf,labelsep=period,justification=raggedright,singlelinecheck=off]{caption}

\makeatletter
\renewcommand{\@biblabel}[1]{\quad#1.}
\makeatother

\date{}

\usepackage{lastpage,fancyhdr,graphicx}
\usepackage{epstopdf}
\begin{document}
\vspace*{0.2in}


\begin{flushleft}
{\Large{}{Quantitative Comparison of Abundance Structures of  Generalized Communities: From B-Cell Receptor Repertoires
to Microbiomes} 
} \\
 Mohammadkarim Saeedghalati\textsuperscript{1}, Farnoush Farahpour\textsuperscript{1},
Bettina Budeus\textsuperscript{1}, Anja Lange\textsuperscript{1},
Astrid M.~Westendorf\textsuperscript{2,3}, Marc Seifert\textsuperscript{4},
Ralf Küppers\textsuperscript{2,4}, Daniel Hoffmann\textsuperscript{1,2,5,6{*}}
\\
 \bigskip{}
\textbf{1} Bioinformatics and Computational Biophysics, Faculty of Biology, University of Duisburg-Essen,
Essen, Germany \\
 \textbf{2} Center for Medical Biotechnology, University of Duisburg-Essen,
Essen, Germany \\
 \textbf{3} Institute of Medical Microbiology, Medical Faculty, University
of Duisburg-Essen, Essen, Germany \\
 \textbf{4} Institute of Cell Biology (Cancer Research), Medical Faculty,
University of Duisburg-Essen, Essen, Germany \\
 \textbf{5} Center for Computational Sciences and Simulation, University
of Duisburg-Essen, Essen, Germany \\
 \textbf{6} Center for Water and Environmental Research, University
of Duisburg-Essen, Essen, Germany \\
 \bigskip{}

\par\end{flushleft}

\begin{flushleft}
%

\par\end{flushleft}

\begin{flushleft}

\par\end{flushleft}

\begin{flushleft}

\par\end{flushleft}

\begin{flushleft}

\par\end{flushleft}

\begin{flushleft}

\par\end{flushleft}

\begin{flushleft}
{*} daniel.hoffmann@uni-due.de
\par\end{flushleft}


\section*{Abstract}

The \emph{community}, the assemblage of organisms
co-existing in a given space and time, has the potential to become
one of the unifying concepts of biology, especially with the advent
of high-throughput sequencing experiments that reveal genetic diversity
exhaustively. In this spirit we show that a tool from community ecology,
the Rank Abundance Distribution (RAD), can be turned by the new MaxRank
normalization method into a generic, expressive descriptor for quantitative
comparison of communities in many areas of biology. To illustrate
the versatility of the method, we analyze RADs from various \emph{generalized 
communities}, i.e.\ assemblages of genetically diverse cells or organisms,
including human B cells, gut microbiomes under antibiotic treatment
and of different ages and countries of origin, and other human and
environmental microbial communities. We show that normalized RADs
enable novel quantitative approaches that help to understand structures
and dynamics of complex generalize communities.


\section*{Author Summary}

Living things are parts of complex communities, similar to humans
living in cities. A quantitative way of describing such communities
is to measure the abundance of each species in the community so that
a sorted list of abundance numbers is produced, a so-called Rank Abundance
Distribution (RAD). With recent breakthroughs in genome analysis this
approach can also be applied to very complex communities, such as
the community of the myriads of microbes in a human gut (gut microbiome),
or the diverse set of human immune cells. One problem with this approach
is that it is not trivial to quantitatively compare RADs for different
communities, especially if they are highly complex. We show that it
is possible to computationally ``normalize'' RADs so that they can
be quantitatively compared across many different communities. In this
way, this normalization enables insight into structures and dynamics
of arbitrary communities. We demonstrate this with applications to
human immune cells, gut microbiomes under antibiotic treatment or
under different nutritional regimes, and environmental microbiomes.



\section*{Introduction}

The \emph{community}, i.e.\ the assemblage of organisms co-existing
in a given space and time, is central to much of ecology \cite{Morin1999CommunityEcology},
and since Darwin's ``entangled bank'' \cite{darwin1859} one of
the great challenges of biology is to explain the observed species
diversity in communities mechanistically as a consequence of interactions
and evolution. Modern experimental methods of high-throughput sequencing
have brought us closer to complete inventories of community diversity.
Moreover, these methods enable us to widen the scope of the community
concept to \emph{generalized communities},
that we define as assemblages of genomically diverse entities, which
include, apart from communities in classical ecology, for instance
B or T cell repertoires of the adaptive immune system, viral quasi-species,
tumors, or human microbiomes.

An intuitive description of a community composition is a table with
columns \emph{species} and \emph{abundance}, possibly ordered from
most to least abundant species (we use the term \emph{species} here
in a loose sense for operational taxonomic units or other genomically
distinct biological entities). A visually more accessible graphical
representation of this table would be a plot that arranges the species
along the horizontal axis and the abundances as vertical bars, sorted
from highest to lowest bar. While such a plot is expressive for a
specific community, it does not lend itself to quantitative comparisons
between communities. To illustrate this point, consider a comparison
of a community of South-American animal species with one of Sub-Saharan
African animal species from regions with otherwise similar conditions.
The two species columns of our table would have practically no overlap
so that a direct comparison of these tables or plots is not possible.
The same lack of overlap has to be expected for other generalized
communities. For instance if we compare high-throughput sequencing
data of B cell receptors of two persons, it is unlikely that there
are receptors on mature B cells that occur in both persons. Nevertheless,
it is a meaningful biological question whether the abundance structures
of the two sets of B cells differ, e.g.\ whether the B cell repertoire
is dominated by a few clones with high cell numbers, or whether it
is distributed over many different clones with low cell numbers.

A popular method for community comparison even in the absence of species
overlap is to compute for each community a diversity index \cite{Magurran2004diversity_book},
i.e.\ a single number that characterizes one aspect of the community,
for instance the species richness, the evenness of the distribution,
the Shannon entropy, or one of many related measures
\cite{Hill:1973}, and then to compare the values of these indices
between communities. The main disadvantage of this index approach
is that it reduces a feature-rich abundance distribution to a single
number, which may neglect important characteristics of that distribution.

An alternative approach that had a major influence on the development
of the theoretical foundations of modern ecology is to discard the
species labels of the species-abundance table, {which
then becomes a so-called Species Abundance Distribution (SAD; for
excellent reviews see \cite{McGill2007}, or \cite{magurran2011biological},
chapter 9). There are several established ways of presenting the information
contained in a SAD (see Fig 1 of \cite{McGill2007}), for instance
as straightforward histogram with species abundance as function of
a species index, or as a binned histogram, typically with doubling
bin widths with decreasing abundance. Here we focus on the Rank Abundance
Distribution or RAD (Fig~\ref{fig1}) as SAD representation. RADs
are simply vectors of species abundances sorted in decreasing order,
usually visualized as two dimensional plots, possibly with one or
both axes scaled logarithmically. In comparison to the mentioned simple
and binned histograms, RADs are more smooth due to their component
sorting \cite{Newman2005}, and they retain the full biological resolution of the sampling
experiment. The information content in RADs and probability distribution
functions is the same, and one can be transformed into the other (see
e.g.~\cite{may1975patterns, Lue2010a}).

Obviously, RADs retain the complete shape information of species-abundance
tables. The abstraction of species information means that RADs enable
analysis of generic abundance distribution features of generalized
communities, independently of the actual species composition.

\begin{figure}[!h]
\includegraphics[width=0.8\textwidth]{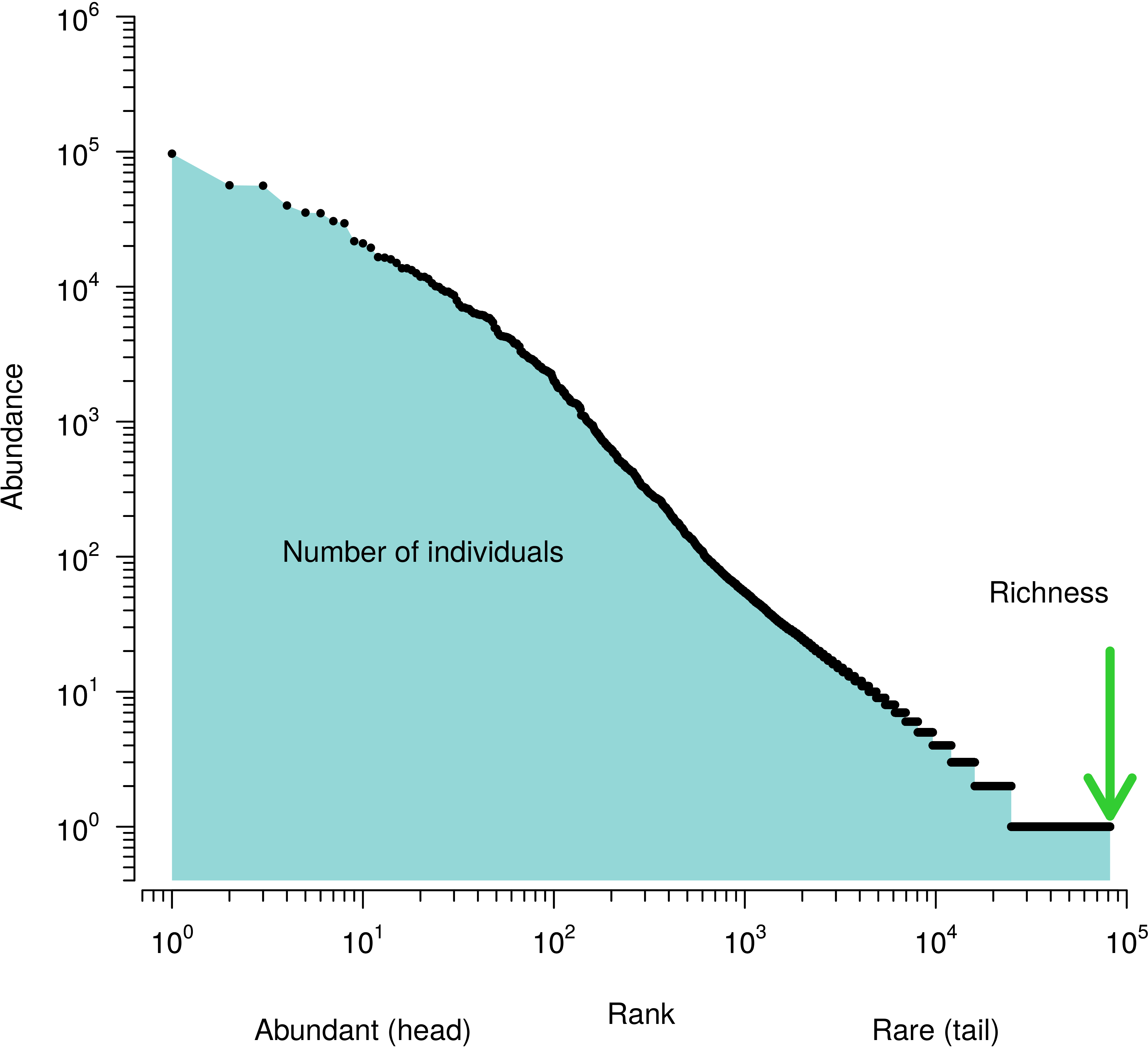}
\caption{\textbf{A typical Rank Abundance Distribution (RAD).} A RAD with species
abundances plotted in decreasing order from the most abundant (rank
$10^{0}=1$) on the left to the least abundant species sampled from
the community on the right. Both axes are scaled logarithmically to
reveal the global structure of the RAD. Quantities such as the number
of sampled individuals or the richness of the sample can be easily
retrieved from the RAD.}

\label{fig1} 
\end{figure}

RADs, and more generally SADs have been a key conceptual tool in the development and benchmarking of mechanistic models of ecological communities \cite{Motomura1932, Fisher1943, Preston1948, MacArthur1957, Hubbell:2001, May2007theoretical}.
The mathematical functions resulting from mechanistic or statistical
models, such as the log-normal distribution, were usually fitted to
empirical RADs or SADs to identify the best community model. This basic research
has paved the way for the application of these distributions to community comparisons,
for instance to the characterization of community changes with changing
environmental conditions (see e.g.~\cite{Gray1979, Foster2010}. In these cases, the
community comparison is typically a parameter comparison between fits
of generic mathematical models to different RADs or SADs, or it is a purely
visual comparison of these distributions \cite{McGill2007}.

In real-world samples, RADs are often not adequately described by
a uniform mathematical model, e.g.\ a single log-normal distribution
\cite{Matthews2015}. In {macroecology}, knowledge
about the properties and relations between the observed animals and
plants can be used to deconstruct multimodal distributions, and to
fit simple models to fractions of the samples \cite{Magurran2004diversity_book}.
This knowledge is generally not available for high-throughput sequencing
data of complex generalized communities, so that
RAD analyses based on simple parametric models are difficult. This
calls for a non-parametric approach. A further problem is that in
practice the number of sampled species or sequences usually differs
between samples. This means that RADs often have different dimensions
and cannot be compared directly. It is possible to test for arbitrary
pairs of RADs the null hypothesis that they originate from the same
distribution using the Kolmogorov-Smirnov test \cite{Tokeshi1993111,Magurran2004diversity_book},
but this is usually not helpful for quantitative comparisons. Technically,
differences between a pair of RADs of different richness $m,n$ could
be also quantified by the Kolmogorov-Smirnov statistic $D$ evaluated
for the corresponding pair of cumulative distribution functions. However,
it is problematic to interpret $D$ in such cases since we are forced
to equate non-reporting of $|m-n|$ ranks in the shorter RAD with
zero abundance, although the non-reporting may have technical reasons,
e.g.\ limited sequencing depth. Thus, the question arises whether
quantitative RAD comparisons are possible between samples of different
richness. {We quote from the widely cited SAD review by McGill \emph{et
al.}~\cite{McGill2007}:}

{``How do we compare SADs? Nearly all comparisons
of SADs along gradients, deconstructions or time trajectories to date
have been purely by visual inspection (...).
Most particularly, these visual inspections have been performed on
rank-abundance plots which, by using an x-axis that runs from 1 to
S (i.e. species richness), seriously confounds the effects of species
richness per se with other changes in the shape of the SAD (...) .
Changes in species richness are a legitimate factor that should be
considered a change in shape of the SAD. However, changes in richness
so strongly dominate in rank-abundance plots that no other changes
are easily considered. Is there any other change in the shape of an
SAD after controlling for the fact that productivity affects richness?
We cannot say at the present time (...) More
rigorous multivariate methods are needed.''}

Here, we introduce MaxRank normalization of RADs, a new method that
enables quantitative comparison of RADs, including their shapes. The
approach non-parametric and allows for the direct quantitative comparison
of complex RADs without deconstruction and model fitting. {An
essential component of the method is the re-sampling of RADs up to
a given richness. Consequently, the resulting normalized RADs (NRADs)
are largely agnostic about the true richness of the original sample.}

The fact MaxRank normalization uses re-sampling may
lead to conflation with \emph{rarefaction} and
\emph{rarefying} (on the distinction
between the terms see \cite{McMurdie:2014}), two other techniques
that also use re-sampling. \emph{Rarefaction}
\cite{Sanders1968} is typically used to estimate and compare richness
between samples, i.e.~exactly the quantity that is not of interest
in MaxRank normalization. \emph{Rarefying } (e.g.~\cite{Koren2013})
is applied to normalize OTU counts between samples. However, it is
precisely the purpose of RADs and MaxRank normalization to abandon
OTUs, and thus to make quantitative comparisons of abundance structures
of different communities possible, irrespective of OTUs.

We show here that results from the quantitative comparison of normalized
RADs reflect biological differences between samples. To emphasize
the versatility of the method, {we have chosen a diverse
set of high-throughput sequencing data representing different types
of generalized} communities, namely, human B cell receptor repertoires,
and various human and environmental microbiomes.

As one example of generalized communities we use
human B cell receptor (BCR) repertoires. The diversity of BCRs in
an individual is crucial for the recognition of antigens and the adaptive
immune response \cite{Janeway:2001}. Here we focus on the so-called
\emph{heavy} part of the receptor encoded by combinations of gene
segments of the Ig heavy chain (IGHV) locus (Fig~\ref{fig2}A; \cite{Tonegawa1983}),
and especially on the diversity of the ``variable'' $V_{H}$ segments
in that part (Fig~\ref{fig2}B). Based on sequence homology, the
$V_{H}$ segments are grouped into seven families ($V_{H}1$ - $V_{H}7$),
with members of a family having more than 80\% sequence homology \cite{Matsuda1996}.
The sizes of the families vary from 1 ($V_{H}6$ family) to 18-21
($V_{H}3$ family). The primary repertoire of rearranged IGHV genes
among na{\"{i}}ve, antigen-inexperienced B cells (Fig~\ref{fig2}C)
typically encompasses all available $V_{H}$ segments, although abundances
can vary considerably between $V_{H}$ gene segments. Exposure to
antigens leads to a selective adaptation of the BCR repertoire that
results in individual-specific sets of memory B cells (Fig~\ref{fig2}D).
In the course of this complex maturation process, the usage of $V_{H}$
segments in BCR rearrangement repertoires may change. On top of this
layer of complexity, several \emph{classes} of BCRs with different
biological functions, such as IgG or IgM, are generated by class switching.
Depending on the chronological order of these different processes,
and on the individual immune histories, we can expect more similar
or more divergent IGHV gene diversity between receptor classes and
individuals in memory B cells. We study this question with RADs computed
from High-Throughput Sequencing (HTSeq) data.

\begin{figure}[!h]
\includegraphics[width=0.8\textwidth]{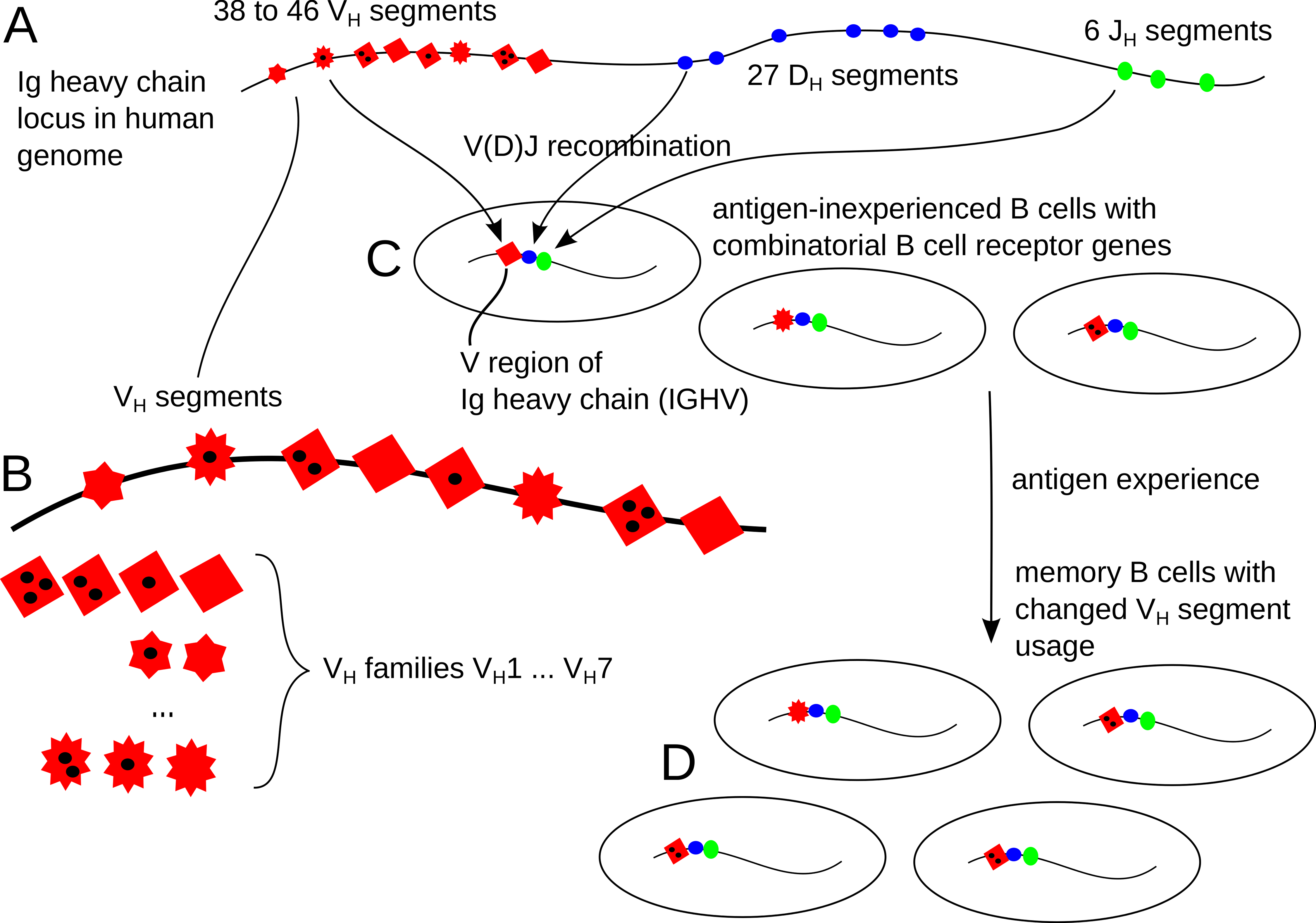}
\caption{\textbf{Diversity of the $V_{H}$ region of BCRs. } (A) The human
genome contains sets of $V_{H}$, $D_{H}$, and $J_{H}$ gene segments.
(B) The ``variable'' $V_{H}$ segments can be grouped into seven
$V_{H}$ families based on sequence similarity. (C) A genetically
diverse pool of B cells is generated by V(D)J recombination. (D) Exposure
to antigens induces an adaptation of the BCR repertoire, generating
genetic variants and changing the usage pattern of $V_{H}$ gene segments.}

\label{fig2} 
\end{figure}

HTSeq technology is also transforming the study of microbial communities,
because it allows us for the first time to see these complex assemblages
in their full diversity \cite{Venter:2004,Tringe:2005}. For instance,
we now start to see the diverse composition of human gut microbiomes,
and we begin to understand the links between the human microbiome,
health and disease \cite{HumanGutMicrobiomConsortium2012a}. However,
the deluge of data makes us also aware of the need for new ways to
analyze and model such complex systems, e.g.\ with methods developed
in ecology \cite{Costello2012a}, such as RADs. We have selected
three HTSeq data sets to demonstrate the potential and limitations
of RADs for the analysis and the modeling of microbiomes: the considerable
effect of antibiotics on gut microbiomes \cite{Dethlefsen2008},
a large collection of gut microbiomes from countries where different
life styles prevail \cite{Yatsunenko2012}, and a diverse set of
human and environmental microbiomes \cite{Caporaso2011e}. In these
examples, we use RADs as an analytic tool to generate easily interpretable
results, and as a basis for quantitative models.

\section*{Materials and Methods}

\subsection*{Datasets}

In the work presented here we used high-throughput sequencing (HTSeq)
amplicon data from four different sources to compute and analyze NRADs,
as described in the following.

\paragraph{B cell dataset}

HTSeq amplicon data of IGHV genes were obtained for four different
memory B cell fractions, $IgG^{+}CD27^{-}$ (for short: $IgG^{+}$),
$IgG^{+}CD27^{+}$, $IgM^{+}IgD^{+}CD27^{+}$, $IgM_{only}^{+}CD27^{+}$,
from two unrelated healthy donors, as published in \cite{Budeus2015a}
and in NCBI Sequence Read Archive (SRA) entry SRP062460. Since each
fraction was split evenly into two independently processed subsamples,
the total number of analyzed samples was 16. Sequence processing was
described in \cite{Budeus2015a}. Briefly, reads of bad quality were
removed, and remaining reads collapsed to single sequences to eliminate
PCR bias. Sequences were then assigned to their respective $V_{H}$
gene segments. For all 16 samples, NRADs were computed using $V_{H}$
gene segments as ``species'' and the number of distinct sequences
assigned to each $V_{H}$ gene segment as abundance of the respective
species.

\paragraph{Gut microbiome dataset}

HTSeq data (bacterial 16S rRNA gene fragment amplicons) from 528 human
gut microbiomes \cite{Yatsunenko2012} were retrieved from MG-RAST
project 401 (http://metagenomics.anl.gov/). The dataset comprised
114 samples from rural Malawi, 315 samples from US metropolitan areas,
and 99 samples from the Amazon region of Venezuela. Human subjects
were aged between 11 days to 83 years with a median of 14 years.

\paragraph{GlobalPatterns dataset}

Caporaso \emph{et al. }\cite{Caporaso2011e} evaluated diversity
patterns of microbial communities across a panel of HTSeq samples
(bacterial 16S rRNA gene fragment amplicons) from diverse sources,
including samples from human feces, skin, and tongue, environmental
samples from ocean, estuary sediment, freshwater, soil, and three
mock communities. Data from 26 of these samples were available through
the R-package phyloseq \cite{McMurdie:2013}, version 1.12.2. We
used the OTU (Operational Taxonomic Unit) tables provided by phyloseq
to compute NRADs. For better readability, samples were renamed according
to the type of sample origin and given a consecutive sample number,
e.g.\ from old names \emph{LMEpi24M}, \emph{SLEpi20M} in \cite{Caporaso2011e}
to new names \emph{lake1}, \emph{lake2}. For human origins \emph{tongue},
\emph{palm}, \emph{feces}, the same sample numbers refer to the same
individuals, e.g.\ \emph{tongue1} and \emph{feces1} come from the
same person number 1. A correspondence table linking old and new names
is provided in \nameref{S1_Table}.


\paragraph{Antibiotic treatment dataset}

Dethlefsen \emph{et al. }\cite{Dethlefsen2008} studied the effect
of the antibiotic Ciprofloxacin on gut microbiomes of three healthy
human individuals using HTSeq data (bacterial 16S rRNA gene fragment
amplicons) before (8 measurements), during (4 measurements), and after
treatment (6 measurements). The corresponding 18 OTU tables were retrieved
from the supplementary material of \cite{Dethlefsen2008}.

\subsection*{Computation of normalized RADs (NRADs)}

For each dataset we followed the flowchart in Fig~\ref{fig3}.

\begin{figure}[!h]
\includegraphics[width=0.8\textwidth]{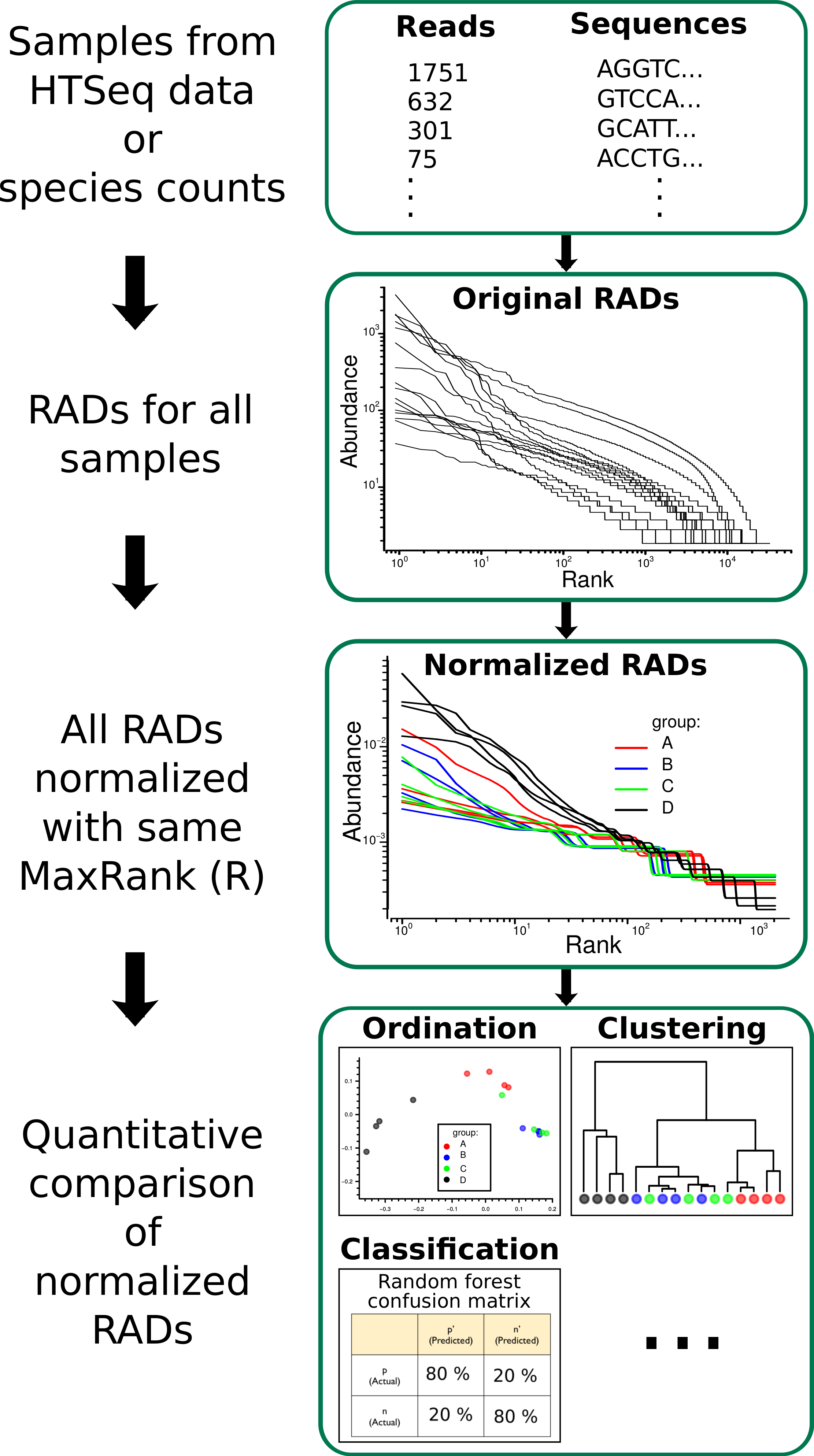}
\caption{\textbf{General process employed in this work.} Flowchart of procedure
from original species/abundances or sequence/reads data (top box)
to original RADs, then to NRADs, and analyses based on NRADs.}

\label{fig3} 
\end{figure}

\paragraph{Compilation of abundance vectors}

For each of the samples we had a two-column table of sequences and
the corresponding numbers of sequence reads (here proxies for abundances),
as obtained from the HTSeq experiment, or, equivalently, a OTU table
with OTU names or numbers and the corresponding abundances in the
sample. All data had been pre-processed and controlled for quality
as described in \cite{Budeus2015a}, \cite{Dethlefsen2008}, \cite{Yatsunenko2012},
and \cite{Caporaso2011e}, respectively.

The species columns were discarded, so that only a list or vectors
of abundances remained for each sample. Abundances of each vector
were sorted in descending order, bringing the highest abundance to
the first element of the vector (= rank 1), the second highest to
rank 2, etc. The resulting abundance vectors, one for each sample,
were then normalized in the next step.

\paragraph{MaxRank normalization}

MaxRank normalization is the key step to make abundance structures
of generalized communities comparable, thus enabling
all succeeding analyses. MaxRank normalization maps all rank abundance
vectors to the same rank range from 1 to a common maximum rank $R$.
The normalization procedure is explained in the following.

First we chose the maximum rank or ``MaxRank'' $R$ (symbol $R$
is used for the MaxRank throughout this work). The minimum $R$ is
2 because $R=1$ would eliminate all abundance structure. The maximum
$R$ is the minimum dimension of rank abundance vectors included in
the analysis. For instance for the GlobalPatterns dataset, we had
rank abundance vectors with dimensions from 2067 to 7679, and the
maximum possible $R$ for the whole set is therefore 2067. A value
of $R$ larger than this maximum would mean that we had to invent
new ranks that have not been observed for at least one sample. In
practice, this maximum $R$ is often a good choice since it retains
the abundance structure of all included communities in the greatest
possible detail. We have therefore chosen in all analyses the maximum
possible $R$.

Once the maximum rank had been fixed to a common $R$ for all samples,
we applied the MaxRank normalization separately to each sample. To
this end we first generated for each sample $s$ a pool of $N_{s}=\sum_{r=1}^{R_{s}}A_{sr}$
individuals, with $R_{s}$ the original maximum rank (i.e.\ the richness)
of sample $s$, and $A_{sr}$ the abundance of rank $r$ in that sample.
From this pool we drew individuals at random with uniform probability
and without replacement as long as the number of sampled ranks did
not exceed $R$. In this way we generated a new, reduced abundance
vector of $R$ ranks, with a reduced number $N_{s}'$ of individuals.
Division of these reduced abundances by $N_{s}'$ transforms the reduced
abundance vector to a probability distribution (or relative abundances) for the $R$ ranks
with rank probabilities summing up to 1. We use therefore the terms
\emph{probability} and \emph{abundance} in the remainder of the article
as synonyms.

If $R=R_{s}$, the procedure above reproduces exactly the original
RAD. If $R<R_{s}$, the random drawing of individuals from the pool
in general introduces a sampling error in the abundances. To control
this error, we repeated the procedure several times (typically 10-100
times) and averaged over all sampled abundance distributions. This
average abundance distribution was returned as the final ``normalized
RAD'' (NRAD) for each sample, together with 90\% confidence intervals
for the mean abundance at each rank, estimated as the interval between the 5\% and 95\% percentile of the bootstrapped averages at the respective rank.

For the NRAD of sample $i$ we use as notation in the following a
vector $\mathbf{a}_{i}$ of $R$ abundances $a_{ir}$: 
\begin{equation}
\mathbf{a}_{i}=(a_{i1},a_{i2},\ldots,a_{iR})\label{eq:notationRAD}
\end{equation}
with elements $a_{ir}\in\,]0,1]$ (ranks $r=1,2,\ldots,R$) that are
sorted ($a_{ik}\ge a_{i\ell}$ for ranks $k<\ell$), and normalized
($\sum_{r=1}^{R}a_{ir}=1$).

\paragraph{Software implementation of MaxRank normalization}

We have implemented the methods used in this work
in free open source software packages \emph{RADanalysis}
and \emph{RankAbundanceDistributions}
available in R (\texttt{https://cran.r-project.org/web/packages/RADanalysis/})
and Julia (\texttt{https://github.com/DanielHoffmann32/RankAbundanceDistributions.jl}).

\subsection*{Analyses of NRADs}

The last box of the flowchart Fig~\ref{fig3} indicates that sets
of RADs normalized to a common $R$ can be analyzed in numerous ways.
In this article we used methods from three branches of data analysis:
ordination, clustering, and classification. Since many ordination
and clustering methods require a \emph{distance} between the studied
objects, we first describe how we computed distances between pairs
of NRADs, and then the actual analysis methods.

\paragraph{Distances between NRADs}

In this work, if not mentioned explicitly otherwise, a distance $d_{R}$
between a pair of NRADs $\mathbf{a}_{i},\mathbf{a}_{j}$ is the Manhattan
distance: 
\begin{equation}
d_{R}(\mathbf{a}_{i},\mathbf{a}_{j})=\sum_{r=1}^{R}|a_{ir}-a_{jr}|.\label{eq:Manhattan}
\end{equation}

The reason for using the Manhattan distance was that it accounts for
NRAD-NRAD differences in a balanced way: NRADs typically show the
largest differences in the first few ranks (in the ``heads'' of
the NRADs, Fig~\ref{fig1}), while the differences are typically
small in the ``tails'' of the NRADs, which comprise many more ranks
than the heads. The Manhattan distance gives the few large differences
in the small heads and the many small differences in the large tails
approximately the same weights.

All pairwise distances $d_{R}(\mathbf{a}_{i},\mathbf{a}_{j})$ between
NRADs in a dataset were collected in a distance matrix for that dataset.
The distance matrix was then used for distance based ordination and
clustering.

We tested the practical suitability of the Kolmogorov-Smirnov statistic
$D$ as a distance measure using the function \emph{ks.test} of R-package
stats \cite{R2015}, version 3.2.2.

\paragraph{Ordination}

We used classical multi-dimensional scaling (cMDS) \cite{Gower1966}
as an ordination method to arrange NRADs (Eq~\ref{eq:notationRAD})
of each dataset in an expressive and visually accessible way, usually
in two dimensions called \emph{first coordinate} and \emph{second
coordinate}. For a cMDS analysis, the distance matrix of the dataset
to be analyzed was submitted to the \emph{cmdscale} function of the
R-package stats \cite{R2015}, version 3.2.2, or the \emph{classical\_MDS}
function of Julia package MultivariateStats, release 0.1.0 (https://github.com/JuliaStats/MultivariateStats.jl).

\paragraph{Clustering}

NRADs of the B cell and GlobalPattern data were clustered hierarchically
by applying function \emph{hclust} of R-package stats, version 3.2.2,
to the NRAD distance matrix using the complete linkage cluster criterion.

\paragraph{NRAD averaging}

Groups of NRADs, for instance the three groups of NRADs of gut microbiomes
of individuals (a) before, (b) during, or (c) after treatment with
Ciprofloxacin \cite{Dethlefsen2008}, or NRADs of gut microbiomes
of individuals in certain age intervals \cite{Yatsunenko2012} were
summarized by computing an average NRAD $\bar{\mathbf{a}}^{(g)}=\frac{1}{n}\sum_{i=1}^{n}\mathbf{a}_{i}$
for each group $g$ with members $i=1,2,\ldots,n$. Note that averages
of NRADs are NRADs themselves with the same MaxRank $R$ and total
abundance of 1. The 90\% confidence interval for the mean NRAD of
each group $g$ was estimated for each rank, estimated as the interval between the 5\% and 95\% percentile of the bootstrapped averages at the respective rank.

\paragraph{Classification}

For the classification of gut microbiomes according to country of
origin we trained random forests \cite{Breiman:2001} models with
R-package randomForest, version 4.6-12 \cite{randomForest2002} with
NRADs as predictors and country (MV vs.\ US) as class labels. The
importance of each rank for the classification was estimated by computing
the effect on the classification performance of randomly permuting
class labels for each rank. Models were tested by threefold cross-validation,
i.e.\ threefold training of a model on a randomly selected 2/3 of
the data, followed by predicting the labels of the left-out 1/3, and
comparison of predictions with ground truth from \cite{Yatsunenko2012}.
Prediction performance was quantified by the accuracy $ACC$ and the
$\kappa$ statistic \cite{Cohen1960kappa}: 
\begin{align}
 & ACC=\frac{n_{correct}}{N},\label{eq:acc}\\
 & \kappa=\frac{ACC-ACC_{expect}}{1-ACC_{expect}},\label{eq:kappa}
\end{align}
with $N$ predictions of which $n_{correct}$ were correct, and $ACC_{expect}$
the accuracy expected by randomly guessing from the given true distribution
of labels with guessing probabilities as obtained from the model.
For example, for $N$ instances labeled by country $X$ or $Y$ we
predict with the model $n_{pred,X}$ and $n_{pred,Y}$, while the
true numbers are $n_{true,X},n_{true,Y}$, and we have then $ACC_{expect}=(n_{pred,X}/N)\cdot(n_{true,X}/N)+(n_{pred,Y}/N)\cdot(n_{true,Y}/N)$.
We report $\kappa$ because $ACC$ can be biased if the different
labels are not represented by approximately equal numbers of instances.
A simplified interpretation of $\kappa$ is the fraction of prediction
accuracy that is not explained by guessing. A perfect model has $\kappa=1$,
a randomly guessing model $\kappa=0$.

\paragraph{Model fitting}

For the B cell data, geometric distributions were fitted to $V_{H}$
gene sequence counts with a maximum likelihood method in Julia v0.4
package \emph{Distributions} (https://github.com/JuliaStats/Distributions.jl)
and tested for consistency by a one-sample Kolmogorov-Smirnov test
in Julia v0.4 package \emph{HypothesisTests} (https://github.com/JuliaStats/HypothesisTests.jl).

For the model of gut microbiome entropy as function of age, we computed
Shannon entropies $H_{R}$ from NRADs with function \emph{entropy}
from Julia v0.4 package \emph{StatsBase} (https://github.com/Julia\-Stats/StatsBase.jl).
Age was provided by the dataset given by \cite{Yatsunenko2012}.

The model was fitted to the set of (entropy, age) pairs by least-squares
minimization with the Levenberg-Marquardt algorithm \cite{Levenberg1944},
as implemented in Julia v0.4 package \emph{LsqFit} (https://github.com/JuliaOpt/LsqFit.jl).
Starting conditions for the fit were the same for the MV and US set,
namely $H_{R}^{0}=3.5$, $\lambda_{R}=0.19$, $H_{R}^{max}=6.0$,
obtained from a rough data-based estimate. Confidence intervals were
estimated from the Jacobians at the optimally fitted parameters, as
implemented in \emph{LsqFit} function \emph{estimate\_errors}.

\paragraph{Comparison to standard distributions}

Five standard distributions commonly used to model
RADs were fitted to the RADs of the GlobalPattern set as described
in \cite{Wilson1991} and implemented in the \emph{radfit} function of R-package vegan, version 2.2-1 \cite{vegan2.2-1}:
broken stick (null model, no free parameter), preemption (geometric
series), log-normal, Zipf, and Mandelbrot. Fits to the Mandelbrot distribution did not converge, so that only the other four distributions are shown.


\section*{Results}

\subsection*{B cell dataset: abundance structures of biologically distinct generalized
communities}

With their highly diverse repertoire of antigen-binding receptors,
human memory B cells are an example of what we have earlier termed
\emph{generalized community}. This diversity
is achieved by a process that is only partly understood and currently
subject of intense research \cite{Choi2013,Hu2015}: it starts with
the genetic recombination of triplets of specific $V_{H}$, $D_{H}$,
and $J_{H}$ gene segments from genetic pools of these segments, followed
by various mutation and selection steps, and eventually leads to distinct
classes and sub-classes of memory B cells.

We assume as a working hypothesis that all memory B cells underwent
the same diversity generating process. If this is true, we should
see a very similar $V_{H}$ gene rearrangement pattern (Fig~\ref{fig2})
in all sub-classes of memory B cells, leading to the same normalized
RADs (NRADs) of memory B cells in all sub-classes. To test this hypothesis,
we used HTSeq data of the immunoglobulin heavy-chain variable (IGHV)
regions of four large memory B cell sub-classes, $IgG^{+}$, $IgG^{+}CD27^{+}$,
$IgM_{only}^{+}CD27^{+}$, $IgM^{+}IgD^{+}CD27^{+}$, from two donors.
The IGHV regions derive from the genomic pool of 38-46 $V_{H}$ segments
(Fig~\ref{fig2}A) and have been modified by mutation and selection
steps. If we interpret the original set of $V_{H}$ segments as the
``species'' to be ranked according to their abundances, we should
under our working hypothesis see the same NRADs in all memory B cell
sub-classes. To exclude distortions of abundances due to primer bias,
we collapsed abundances of $V_{H}$ segments from the measured read
numbers to the numbers of distinct $V_{H}$ sequence variants. Thus,
in this case an abundance is the number of distinct sequences that
originate from the same $V_{H}$ segment, and that have been diversified
by somatic mutations and clonal expansions. Fig~\ref{fig4} summarizes
the results.

\begin{figure}[!h]
\includegraphics[width=0.8\textwidth]{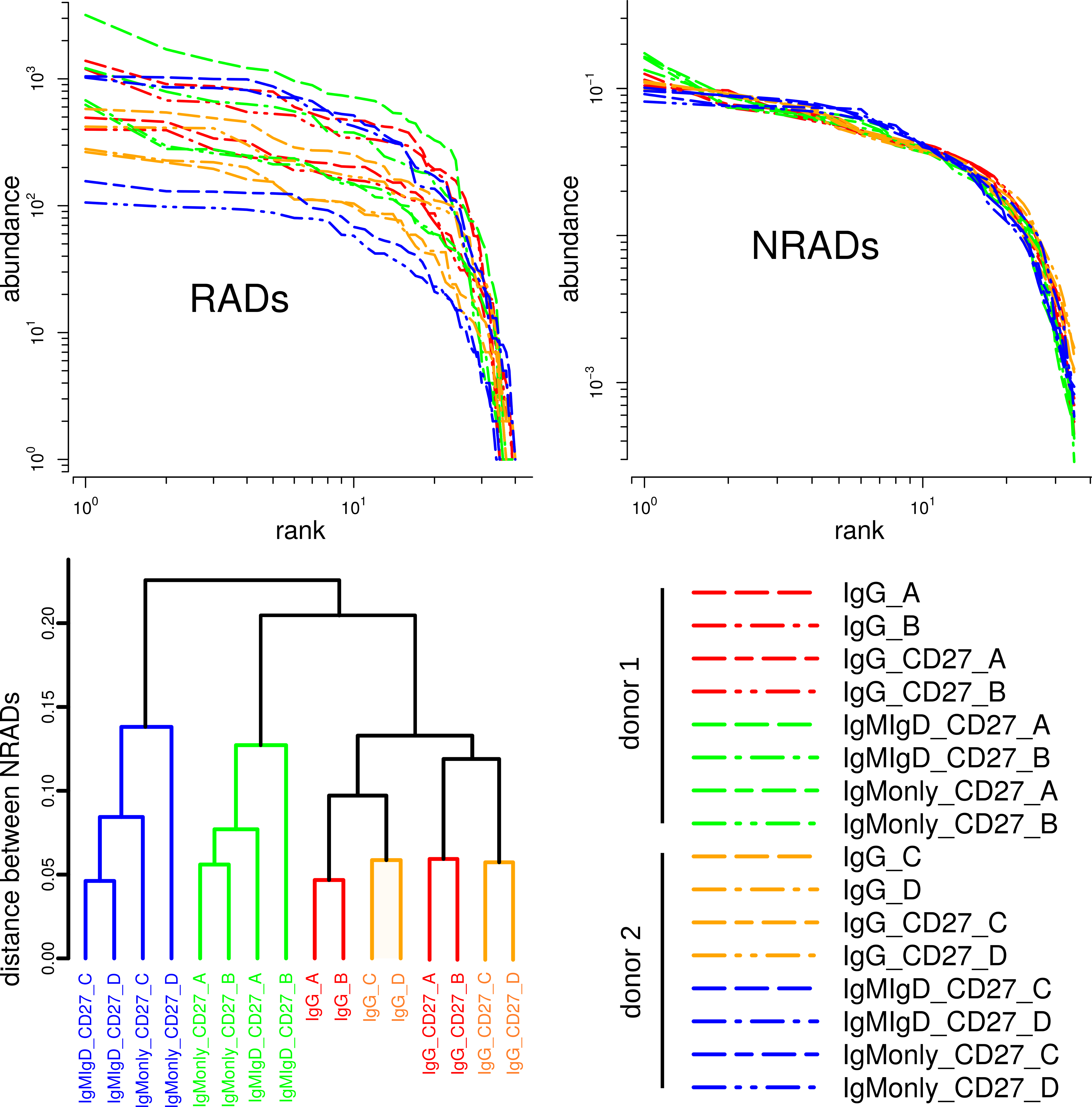}
\caption{\textbf{Rank abundance distributions of memory B cell receptors}.
Four different B cell receptor sub-classes from donors 1 (replicate
samples A, B) and 2 (replicates C, D) are compared. Top left panel:
Log-log plot of RADs prior to normalization. Top right: Log-log plot
of corresponding NRADs. Legend for RADs and NRADs is given in bottom
right panel. Bottom left: Hierarchical clustering tree based on all
pairwise distances between the 16 NRADs.}

\label{fig4} 
\end{figure}

The non-normalized RADs (top left of Fig~\ref{fig4}) have similar,
boomerang-like shapes, although direct comparisons is difficult since
differences in abundance span more than one order of magnitude, and
maximum ranks differ between 35 and 40. For direct comparison we therefore
normalized the RADs to a MaxRank $R=35$ (top right of Fig~\ref{fig4}).
The resulting NRADs have overall very similar shapes, lending support
to our working hypothesis of a common generation and selection process.
However, there are notable features that differentiate between groups
of NRADs. For instance at rank 1 the most diverse IGHV regions in
IgM receptors of donor 1 (green curves in Fig~\ref{fig4}) are more
abundant than all other rank 1 abundances. Conversely, for donor 2
the most diverse IGHV regions in IgM (blue curves) are the least abundant
of all rank 1 abundances. Towards higher ranks, IgM abundances of
both donors (blue and green) are more similar to each other. IgG receptors
(red and orange) have more similar abundance structures throughout
all ranks, and have stronger right tails than IgM receptors, indicating
a more even $V_{H}$ segment diversity in IgG than IgM receptors.

The differences between the NRAD curves are subtle, but they emerge
clearly when we quantitatively analyze NRAD distances (Eq~\ref{eq:Manhattan}).
In the hierarchical clustering tree of the distances (Fig~\ref{fig4})
we see three main clusters of NRADs, a big cluster of $IgG^{+}$ and
$IgG^{+}CD27^{+}$ NRADs on the right of the tree in red and orange,
and two clusters of IgM related NRADs. These three clusters appear
robustly, no matter whether the average linkage or the complete linkage
criterion is used for clustering.

As expected for replicates, (A, B) and (C, D) of the same sub-class
coming from the same donor yield NRADs that are most similar and thus
fall into the same lowest-level clusters. Beyond this, memory B cell
sub-classes have remarkably different cluster structures. The big
$IgG^{+}$/$IgG^{+}CD27^{+}$ cluster (red and orange) of eight NRADs
has a substructure of an $IgG^{+}$ cluster and a separate $IgG^{+}CD27^{+}$
cluster, i.e.\ here the memory B cell sub-class has a stronger impact
on the NRAD than inter-donor differences. This is different for the
$IgM_{only}^{+}CD27^{+}$ and $IgM^{+}IgD^{+}CD27^{+}$ clusters.
There, each of the two donors forms a cluster of its own that combines
both $IgM_{only}^{+}CD27^{+}$ and $IgM^{+}IgD^{+}CD27^{+}$ NRADs,
i.e.\ in these two IgM sub-classes, inter-personal differences in
$V_{H}$ diversity patterns are stronger than differences between
sub-classes.

Our working hypothesis was that we have basically a single, diversity
generating process for all memory B cell sub-classes, leading to the
same NRADs for $V_{H}$ gene rearrangement pattern in all sub-classes.
Overall, our results are consistent with this big picture since all
NRADs are variants of the same boomerang shaped template. However,
the quantitative analysis of NRAD distances picked up differences
that require refinements of this model. This is not surprising since
some of the steps that potentially affect $V_{H}$ rearrangement diversity,
for instance class switching, cannot apply equally to all memory B
cell sub-classes. What is surprising are the specific differences
in $V_{H}$ rearrangement diversity between IgM and IgG sub-classes,
e.g.\ the structure of the IgG cluster discussed above suggests that
there could be significantly different selection pressures towards
the final memory B cells in the two sub-classes $IgG^{+}$ (i.e.\ $IgG^{+}CD27^{-}$)
and $IgG^{+}CD27^{+}$.

Recently, we have used the same HTSeq data for a detailed sequence-based
analysis of the clonal composition and genealogy of memory B cells
\cite{Budeus2015a}. Although our current NRAD-based analysis disregards
much of the information used in \cite{Budeus2015a}, results are
consistent: First, the $V_{H}$ gene rearrangement composition is
mostly very similar across the studied sub-classes of memory B cells,
in agreement with a shared generation process \cite{Seifert2009}.
Second, $IgM^{+}IgD^{+}CD27^{+}$ and $IgM_{only}^{+}CD27^{+}$ show
almost the same $V_{H}$ gene rearrangement diversity, likely due
to their clonal relatedness as described in \cite{Budeus2015a}.
Third, there are significant and consistent differences between $IgG^{+}CD27^{+}$
and $IgG^{+}CD27^{-}$ with respect to mutation load in both donors,
also in agreement with \cite{Fecteau2006} or \cite{Jackson2014}. 

To conclude this section, we return to the conspicuous boomerang shape
that is the template common to all B cell receptor RADs and NRADs
(Fig~\ref{fig4}). When testing for similarity to standard model
distributions in ecology, we found that the broken stick distribution
\cite{MacArthur1957} is a good description for the NRADs of sub-class
$IgG^{+}CD27^{+}$ (Fig~\ref{fig5}). If included in the hierarchical
clustering, the broken stick NRAD appears among the branches of the
$IgG^{+}CD27^{+}$ sub-tree (inset of Fig~\ref{fig5}). However,
even for the other sub-classes, we cannot reject the broken stick
distribution (p-values from Kolmogorov-Smirnov tests between 1.0 and
0.87 with a median of 0.99), though they deviate more than $IgG^{+}CD27^{+}$.
Note that the normalized broken stick distribution in Fig~\ref{fig5}
has no free parameters and therefore has not been fitted.

\begin{figure}[!h]
\includegraphics[width=0.8\textwidth]{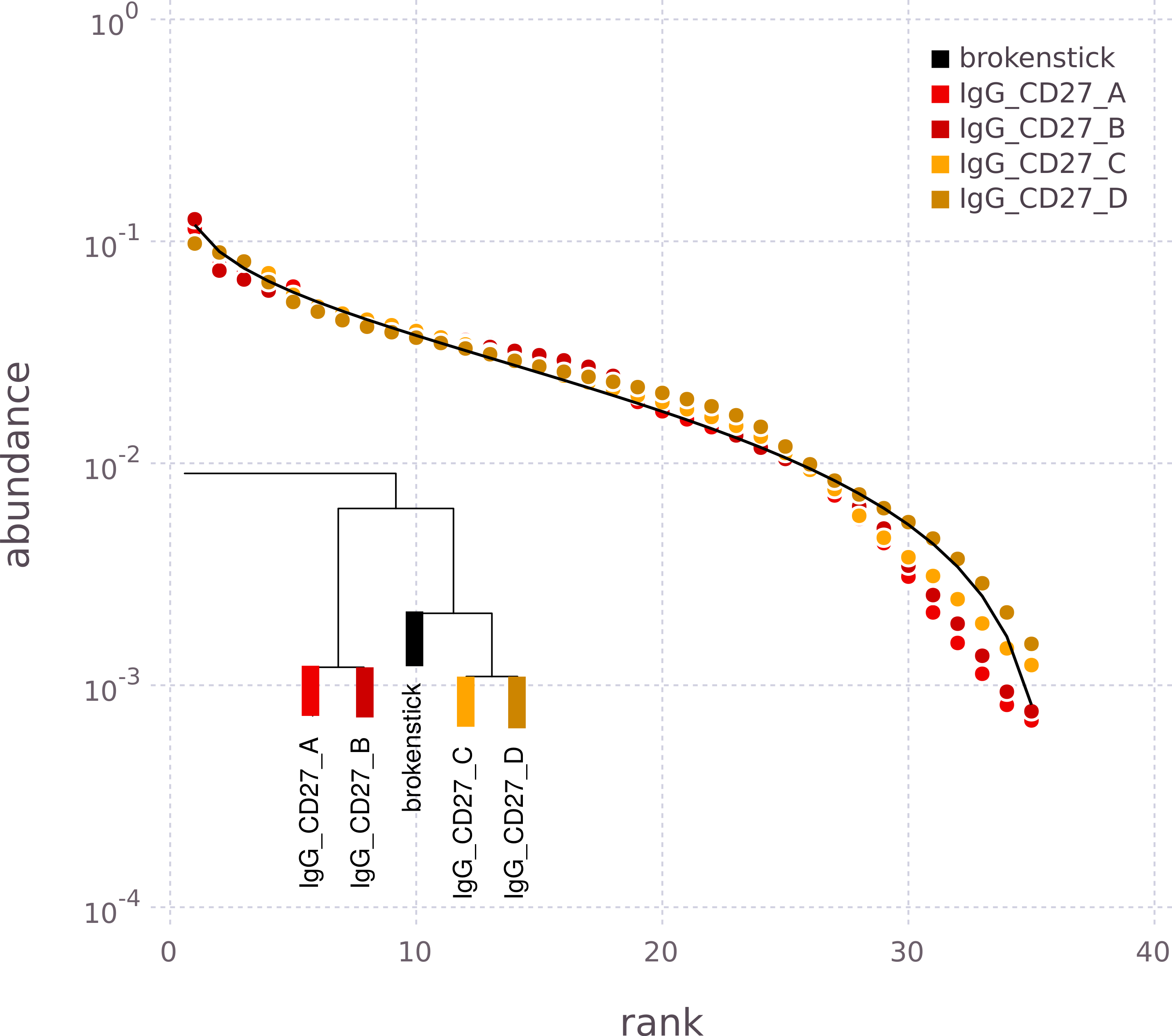}
\caption{\textbf{Broken stick distribution (solid line) and NRADs of $IgG^{+}CD27^{+}$
fractions (points).} Inset: section of hierarchical clustering dendrogram
where broken stick distribution appears. This plot adopts the usual
presentation of the broken stick distribution in the literature with
linear horizontal axis and logarithmic vertical axis. Therefore the
boomerang shapes of the log-log Fig~\ref{fig4} appear horizontally
stretched.}

\label{fig5} 
\end{figure}

Several mechanisms in community ecology and elsewhere lead to broken
stick RADs \cite{MacArthur1957,Cohen1968,Sugihara1980,Tokeshi1990}.
A simple explanation for the observed RADs of $V_{H}$ segment usage
could be the following. Assume a fixed number $n_{V_{H}}$ of $V_{H}$
segments in the genome, and a fixed total number $N_{t}$ of all BCR
sequence variants (i.e.\ summed over all $V_{H}$ segments). The
biological purpose of fixing $N_{t}$ could be to provide a sufficient
number of BCR variants to cover the typical antigen diversity. These
two assumptions fix the average number $N_{t}/n_{V_{H}}$ of sequence
variants per $V_{H}$ segment. If this is all we know, the Maximum
Entropy principle states that the geometric distribution is the most
parsimonious explanation fulfilling these requirements \cite{Cover2006ElemInformTheo}.
The geometric distribution is the discrete equivalent of the continuous
exponential distribution, which generates the broken stick RAD \cite{Cohen1968}.
Thus, our argument posits a random process that produces numbers of
sequence variants per $V_{H}$ segment with a geometric distribution.
In fact, sequence counts of most RADs are compatible with geometric
distributions according to Kolmogorov-Smirnov tests at significance
level 0.05 (exception: sample D of $IgM^{+}IgD^{+}CD27^{+}$ with
$p=0.04$). Given that there is good agreement between the RADs of
the BCR sub-classes (Fig~\ref{fig4}), and also good agreement in
the usage of individual $V_{H}$ segments between human donors \cite{Budeus2015a},
our argument suggests the following two testable hypotheses. First,
the random mechanism leading to the broken stick RAD could be encoded
in the human genome and conserved among individuals with intact immune
systems. Second, since our argument is generic, we should see the
broken stick RAD also in other species with similar BCR rearrangement
mechanisms.

\subsection*{Antibiotic treatment dataset: abundance structure reacts to perturbations}

Dethlefsen \emph{et al.~}\cite{Dethlefsen2008} reported the effects
of a short course of Ciprofloxacin (Cp) treatment on the gut microbiomes
of three healthy human individuals. When comparing gut microbiomes
prior to treatment and during treatment, they found markedly perturbed
taxonomic composition, richness, diversity, and evenness. These perturbations
varied between individuals. After treatment, the community compositions
recovered within four weeks to states close to pre-treatment, though
with some species lost. We tested whether the dynamics of perturbation
and recovery is reflected by changes of NRADs.

The NRADs fall into two clusters, one well-defined ``off-Cp'' cluster
of NRADs before and after treatment (blue and green in Fig~\ref{fig6}),
and one wider ``on-Cp'' cluster during treatment (red in Fig~\ref{fig6}).
All on-CP NRADs have heavier heads and less weight in the tails, consistent
with the decreased gut microbiome diversity under treatment discovered
by \cite{Dethlefsen2008}. After normalization, we could compute
distances between all pairs of NRADs and apply multidimensional scaling
(MDS) to the distance matrix. The MDS plot Fig~\ref{fig6}B captures
the abundance dynamics from the well-defined off-Cp cluster on the
right to the wider on-Cp cluster on the left and back again to the
off-Cp cluster on the right. We usually find in MDS analyses a strong
correlation of the first coordinate with Shannon entropy. Hence, the
dynamics in the MDS plot Fig~\ref{fig6}B from right (pre-Cp) to
left (Cp) to right (post-Cp) corresponds to a succession of high-low-high
entropy. This can also be seen directly from the NRADs: the off-Cp
NRADs have a relatively heavy tail and weak head, i.e.\ a more even
distribution with higher entropy, corresponding to a more diverse
gut microbiome. Conversely, the on-Cp NRADs have a more heavy head
and weaker tail, i.e.\ a less even distribution with lower entropy,
corresponding to a less diverse gut microbiome, partly decimated by
the effect of the antibiotic.

\begin{figure}[!h]
\includegraphics[width=0.8\textwidth]{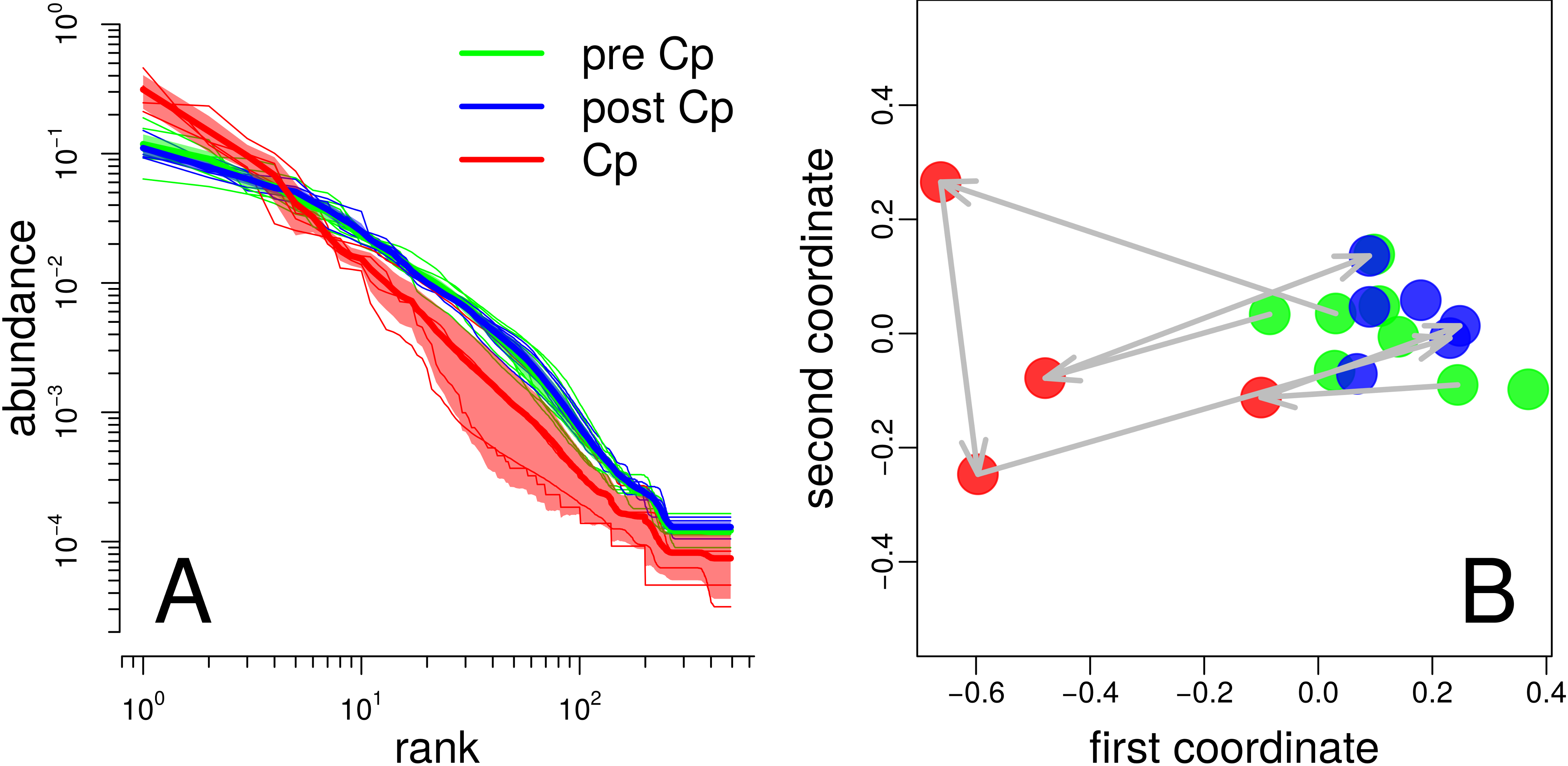}
\caption{\textbf{Abundance dynamics of gut microbiomes of three individuals
under treatment with antibiotic Ciprofloxacin (Cp).} (A) NRADs before
(green), during (red), and after (blue) treatment. Bold lines are
mean NRADs, shaded regions are 90\% confidence intervals of the means.
(B) MDS of NRADs with one point per NRAD using the same color code
as in panel A. For each of the three individuals, arrows connect points
corresponding to the last measurement before treatment, measurements
during treatment, and the first measurement after treatment. The
two coordinates of the MDS plot explain 89\% of the NRAD distances.}

\label{fig6} 
\end{figure}

Dethlefsen \emph{et al.~}\cite{Dethlefsen2008} remarked that after
treatment several taxa failed to recover, while the participants in
the study had normal intestinal function, and they argued that the
eliminated taxa after treatment may have been replaced by other taxa
with similar functions. The fact that pre- and post-Cp NRAD ensembles
have the same shapes and form a single compact off-Cp cluster (Fig~\ref{fig6})
supports this assessment.

It is instructive to compare our analysis based on the abundance structure with an OTU composition analysis as in Fig~6 of Ref~\cite{Dethlefsen2008}. The OTU based PCA in Ref~\cite{Dethlefsen2008} has a cluster structure that is influenced by both the individual microbiome donor and by the treatment state. The conflation of both influences makes the result of the PCA richer but also more difficult to interpret: If we consider individual microbiome compositions, all three individuals have different microbiome dynamics under treatment. Conversely, the NRAD based analysis is blind to individual differences in microbiome composition. This blindness to composition means on the other hand to focus on the abundance structure, which makes the result in our Fig~6 more easy to interpret: In terms of the abundance structure, all three individuals behave in the same way, clearly showing a generic effect of the antibiotic treatment.

\subsection*{Gut microbiomes dataset: NRADs enable quantitative models}

Yatsunenko \emph{et al.~}\cite{Yatsunenko2012} found in 528 gut
microbiomes from Malawi, United States and Venezuela, that (1) species
richness gut microbiomes increased with age from birth to about the
third year, and then was much less variable, and (2) taxonomic composition
of adult gut microbiomes from the Unites States differed strongly
from those of Malawi and Venezuela, while the latter two showed less
pronounced differences. In our analysis we do neither use richness
(we normalize to a common richness), nor taxonomic labels, but we
use solely the abundance vectors (Eq~\ref{eq:notationRAD}) as quantitative
descriptors of NRAD shapes. Nevertheless, we will in the following
show results consistent with key results from \cite{Yatsunenko2012}
with NRADs. Additionally, we will present a novel NRAD-based quantitative
model for the development of gut microbiome entropy as function of
age.

Prior to normalization, the richness of the samples covered three
orders of magnitude, from $4105$ to $296214$ different ranks. All
$528$ RADs were normalized to the same MaxRank of $R=4105$ to make
RAD shapes and quantities computed thereof comparable.

\paragraph{NRADs differentiate between ages and countries}

First we applied MDS to the distance matrix of all 528 NRADs to see
whether NRADs of different countries (especially between US = United
States and MV = Malawi/Venezuela) and ages can be distinguished (Fig~\ref{fig7}A).
We found a banana shaped distribution in the MDS plot, arranged along
the first coordinate that explains two thirds of the spread between
NRADs (the banana shape also appears with nonmetric
MDS (not shown) \cite{gardener2014community}). The banana reaches from babies and
small children on the left to adults on the right. Thus we have a
clear age related trend of NRADs. It is remarkable that while the
points for the smallest children have the largest scatter, the averages
between MV and US are the same within the margin of error. Above the
youngest age group, NRADs split up into two branches, one for MV and
one for US. For adults the means of the MDS clusters of MV and US
have small errors and are clearly separated. Differences between the
older age groups within the same country are small. The described
patterns in the MDS plot is consistent with the taxonomy based results
in \cite{Yatsunenko2012}.

\begin{figure}[!h]
\includegraphics[width=0.8\textwidth]{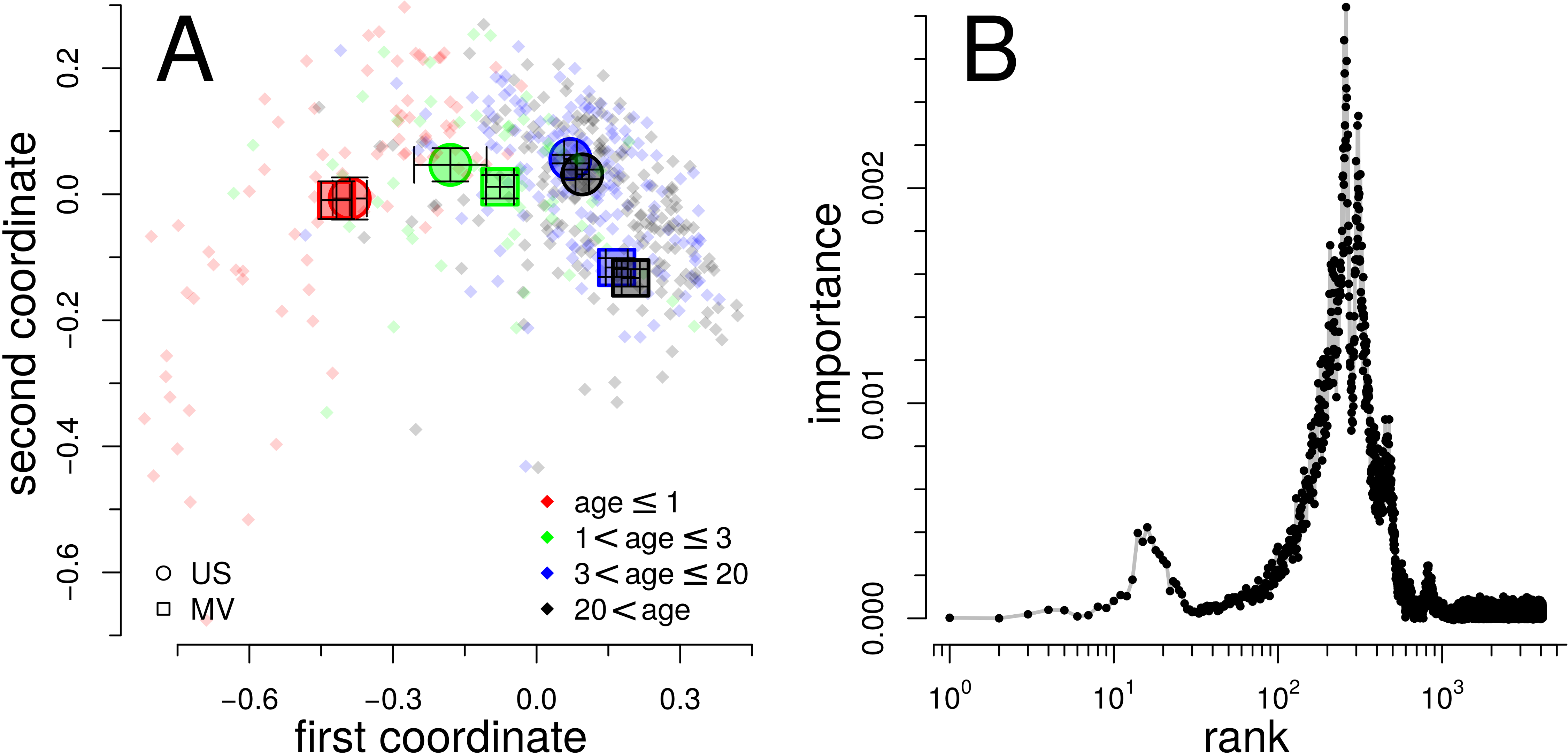}
\caption{\textbf{Country of origin and age as determinants of gut microbiomes
NRADs.} (A) MDS-ordination of NRADs of those 489 gut microbiomes from
Malawi/Venezuela (MV) and United States (US) with age information.
Small symbols represent individual NRADs, large symbols are averages.
Error bars are 90\% confidence intervals of the averages. {The
two coordinates of the MDS plot explain 83\% of the NRAD distances.
}(B) Importance of each of the 4105 NRAD ranks for the random forest
classification according to country of origin (MV vs.\ US). The two
peaks around ranks 20 and 200 are the NRAD regions that carry most
information about the country of origin.}

\label{fig7} 
\end{figure}

Yatsunenko \emph{et al.~}\cite{Yatsunenko2012} had trained a statistical
model that identified bacterial species characteristic to each of
the countries. These species could be used to predict the country
of origin from the taxonomic composition.

The pattern in the MDS plot (Fig~\ref{fig7} A) suggests that it
could be possible to train for the older age groups a statistical
model that correctly predicts the country of origin of a sample from
the shape of the NRAD. To test this hypothesis we trained a random
forest model to classify NRADs from individuals older than 3 years
according to country Malawi/Venezuela (MV, 89 NRADs) or United States
(US, 254 NRADs). We found a high accuracy $ACC=0.94\pm0.02$ (mean
$\pm$ standard error) of the model in threefold cross-validation.
However, since the dataset is by far not evenly distributed over both
countries, $ACC$ could grossly overrate the performance. We therefore
computed the $\kappa$-statistic and found $\kappa=0.85\pm0.05$,
confirming the very good performance of this statistical model for
predicting country of origin from NRAD. To avoid misunderstandings,
we emphasize that ``country of origin'' is here a proxy for conditions,
e.g.\ life style or diet, prevailing in a sample set that lead to
a certain type of NRAD. If these conditions are similar, it is conceivable
that NRADs will be similar too and therefore cannot be separated accurately.

It is interesting that not the head or tail region is most important
for the classification performance, but that two separate regions
in the middle carry most of the information about the country of origin
(Fig~\ref{fig7}B). Thus we could have normalized to even smaller
$R$ values without too much loss of information for the classification.

\paragraph{A quantitative model for the change of gut microbiome NRAD entropies
with age}

Yatsunenko \emph{et al.~}\cite{Yatsunenko2012} observed that the
taxonomic richness in gut microbiomes increased strongly during the
first 3 years of age, and then stabilized. We asked whether NRADs
reflect this dynamics, even though NRADs ignore taxonomy and eliminate
richness. For the first analysis (Fig~\ref{fig8}) we split the dataset
into log-age intervals of approximately equal lengths, so that we
have shorter age intervals at young age when most of the changes are
expected, and longer age intervals in the more stable later regime.
We then averaged all NRADs in these intervals, irrespective of country
of origin.

\begin{figure}[!h]
\includegraphics[width=0.8\textwidth]{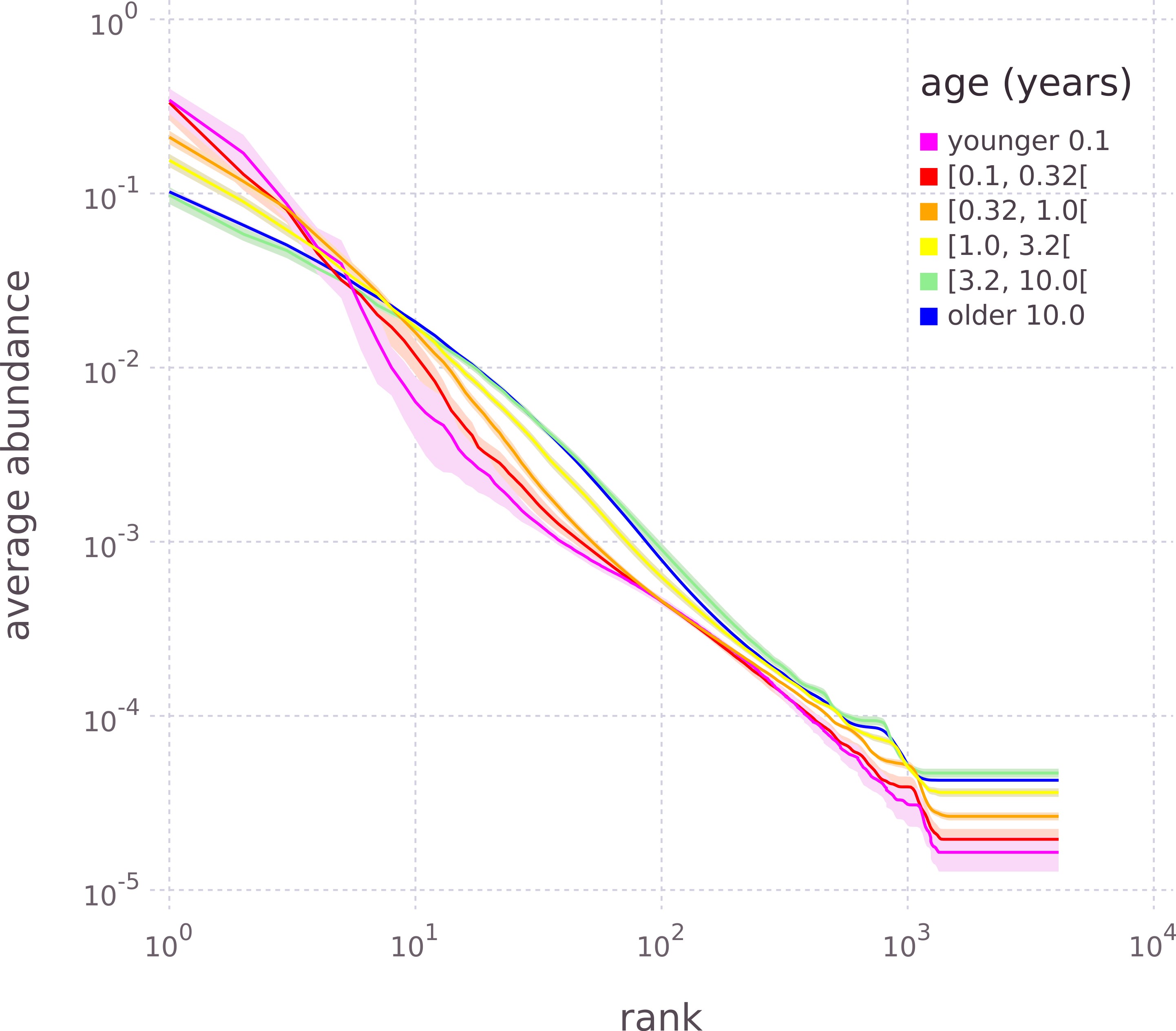}
\caption{\textbf{Averaged NRADs of gut microbiome data in six age groups.}
The number of NRADs per group from youngest to oldest were 9, 18,
55, 64, 34, and 309, respectively. Solid lines are mean NRADs, shaded
areas are 90\% confidence intervals for the means.}

\label{fig8} 
\end{figure}

The NRADs in Fig~\ref{fig8} have a clear dynamics with age: The
average abundance in the head region up to about rank $10^{0.6}\approx4$
decreases continuously from the youngest to the oldest age groups,
while in the middle and tail regions the average abundance increases
from youngest to oldest. These changes are fast in the youngest age
groups, and slow down in the oldest groups that have practically the
same NRADs. The change from NRADs with strong heads and weak tails
in the young, to NRADs with weaker heads and stronger tails in the
old means that the evenness of the distribution increases. Since evenness
is one aspect of diversity \cite{Magurran2004diversity_book}, we
conclude that the growing species richness described by \cite{Yatsunenko2012}
in their taxonomy-driven analysis is mirrored by an increase of evenness
with age as computed from NRADs.

We now proceed from the qualitative comparison of NRADs in Fig~\ref{fig8}
to a quantitative model of Shannon entropy as a function of age. Quantitatively,
Shannon evenness $J^{(i)}$ of NRAD $i$ can be expressed in terms
of entropy $H^{(i)}$ and species richness $S^{(i)}$ as \cite{Magurran2004diversity_book}:
\begin{equation}
J^{(i)}=\frac{1}{\log S^{(i)}}\cdot H^{(i)}\mathrm{\;\;\; with\;\;\;}H^{(i)}=-\sum_{r=1}^{R}a_{ir}\log a_{ir},\label{eq:Shannon}
\end{equation}
where we have used rank abundances $a_{ir}$ according to Eq~(\ref{eq:notationRAD}),
and, for simplicity, plain symbols $J$, $H$ instead of the commonly
used $J'$, $H'$. Since by definition for MaxRank normalized RADs
we have $S=R=\mathrm{const}$, the evenness $J$ computed from NRADs
is proportional to the Shannon entropy $H$ of these NRADs. Thus,
we can rephrase our observation that NRAD evenness $J$ grows with
age as an increase of NRAD entropy $H$ with age, or, because entropy
is a measure of diversity, as growth of diversity with age.

Entropy is sensitive to sampling errors, and therefore not an ideal
measure of diversity \cite{Magurran2004diversity_book}. For instance,
bigger samples from the same biological system have often higher richness
which directly affects entropy. MaxRank normalization eliminates richness
and thus attenuates this error. This property invites quantitative
comparison of evenness or entropy across many samples. Here we exploit
this to study the 489 HTSeq datasets of gut microbiomes that had information
about age in their metadata. The richness in these samples varies
over three orders of magnitude, between $\num{4e3}$ and $\num{3e5}$.

In Fig~\ref{fig9}A entropies of the 489 gut microbiomes with age
information are plotted against age. The dynamics of average entropy
with age is consistent with Fig~\ref{fig8} discussed previously,
namely a strong increase in the youngest and stabilization in older
individuals.

\begin{figure}[!h]
\includegraphics[width=0.8\textwidth]{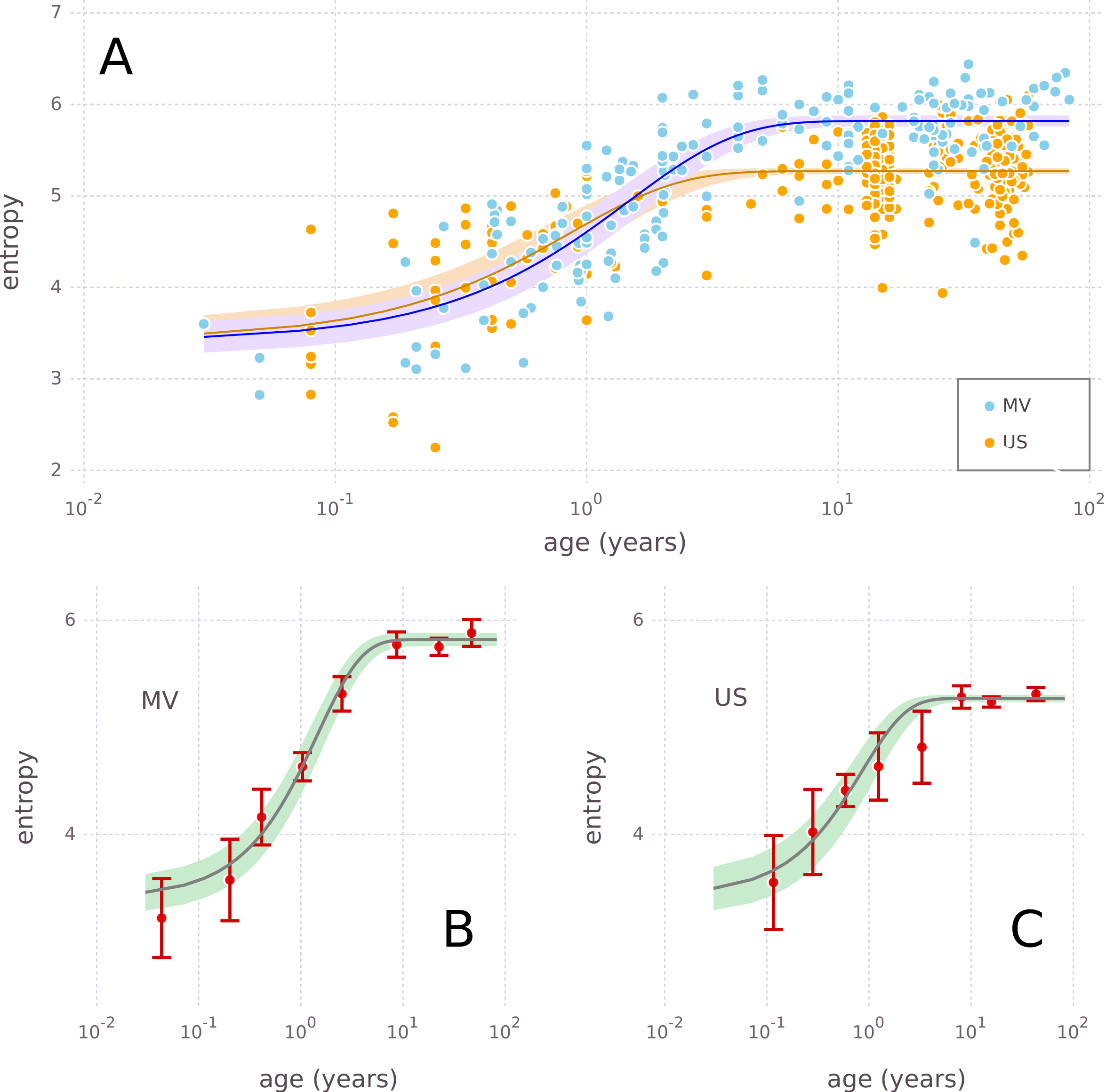}
\caption{\textbf{Development of gut microbiome entropy $H_{R}$ with age $t$.}
(A) Entropies $H_{R}^{(i)}$ (with $R=4105$) in nats for the 213
samples from Malawi and Venezuela (MV, blue dots), and the 315 samples
from the United States (US, orange dots). Log-scaled horizontal axis
is age in years. Superimposed are models for $H_{R}^{MV}(t)$ (blue
line) and $H_{R}^{US}(t)$ (orange line) according to Eq~(\ref{eq:H_of_t}).
Areas around the model lines shaded in blue and orange are the corresponding
90\% confidence intervals of the respective models. (B) and (C) Comparison
of mean entropies of measured data (red points) and their corresponding
90\% confidence intervals (error bars), with the model (solid gray
lines) and its 90\% confidence interval (shaded areas), for MV (panel
B) and US (panel C). Model lines and shaded areas are the same as
in panel A.}

\label{fig9} 
\end{figure}

The empirically observed change of entropy with age could be explained
by a simple quantitative model. We assume that the youngest babies
have a gut microbiome of very low diversity. Every microbial intake
by the child will therefore have potentially a large impact on the
diversity of its microbiome. This will lead to an increasing diversity
of the gut microbiome with age. However, as the diversity increases,
the impact of new intakes on the diversity will decrease since some
of the species have already been taken up earlier. Thus we expect
an increase that asymptotically approaches a diversity typical for
the environment and life style of the individual. One of the simplest
models for entropy $H_{R}$ as measure of diversity with age $t$
that shows this behavior is: 
\begin{equation}
H_{R}(t)=H_{R}^{max}-(H_{R}^{max}-H_{R}^{0})\, e^{-\lambda_{R}t},\label{eq:H_of_t}
\end{equation}
with three parameters, the maximum entropy $H_{R}^{max}$ defining
the asymptotic diversity, the entropy $H_{R}^{0}$ of the gut microbiome
shortly after birth, and the entropy growth rate $\lambda_{R}$. All
quantities have an index $R$ to remind us that we base our model
on NRADs with a certain MaxRank $R$.

We have fitted two sets of optimal parameters, one with the MV and
one with the US data (Table~\ref{tab:entropy-age-parms}).

\begin{table}[!ht]
\centering \caption{ \textbf{Optimal parameters for the model of gut microbiome NRAD entropy
as function of age.}}

\begin{tabular}{|l|l|l|l|}
\hline 
 & {\boldmath $H_{R}^{0}$}  & {\boldmath $H_{R}^{max}$}  & {\boldmath $\lambda_{R}\;(yr^{-1})$}\tabularnewline
\hline 
$MV$  & $3.41\pm0.17$  & $5.82\pm0.06$  & $0.69\pm0.09$ \tabularnewline
\hline 
$US$  & $3.43\pm0.19$  & $5.27\pm0.03$  & $1.16\pm0.27$ \tabularnewline
\hline 
\end{tabular}

\begin{raggedright}
Parameters of optimal models Eq~(\ref{eq:H_of_t}) fitted separately
to the data from MV and US. Errors are 90\% confidence intervals. 
\par\end{raggedright}

\label{tab:entropy-age-parms} 
\end{table}

{\boldmath $H_{R}^{0}$} $H_{R}^{0}$ The corresponding models
are the solid lines in Fig~\ref{fig9}. Fig~\ref{fig9}B and C show
that the fitted models capture the average development of diversity
with age and the differences between MV and US. The models have a
reasonable accuracy with coefficients of determination of $r_{MV}^{2}=0.74$
and $r_{US}^{2}=0.52$. For all points in Figs~\ref{fig9}B and C
data and model are consistent, with one boundary case being the US
model around age $3~yr$ ($\approx10^{0.5}~yr$), where it has a larger
error because only a small number of measurements were available (see
Fig~\ref{fig9}A at this age).

The models are consistent with the observations by \cite{Yatsunenko2012}
in their taxonomy-based analysis, such as the increase of gut microbiome
diversity in babies and children up to about three years, the split
between MV and US in small children, and the turn to the higher asymptotic
level in MV and the lower asymptotic level in US. Quantitatively,
the model predicts that the youngest babies on average have gut microbiomes
with $H_{4105}\approx3.42$ in both MV and US. The averages cannot
be distinguished between MV and US up to about $2~yr$ ($=10^{0.3}~yr$).
On average, children in the US reach the plateau $H_{4105}^{max,US}=5.27$
at about $3.2~yr$ ($=10^{0.5}~yr$), children in MV reach the higher
plateau $H_{4105}^{max,MV}=5.82$ at about $5.6~yr$ ($=10^{0.75}~yr$).
It must be emphasized that the model predicts \emph{average} diversities
as a function of age, not individual diversities. The large spread
of entropy values around the average model in Fig~\ref{fig9}A shows
that inter-individual variation cannot be neglected. Finally, although
the model is simple, plausible and fits the data well, other mathematical
forms than Eq~(\ref{eq:H_of_t}) are possible. 






\subsection*{GlobalPatterns dataset: strengths and limitations of NRADs}

Analyses with NRADs are complementary to taxonomy-based analyses:
NRADs are blind to taxonomy, which is both a limitation and an advantage.
It is a limitation because community biology is a function of taxonomic
composition. It is an advantage because it enables quantitative comparison
between taxonomically different generalized communities
and thus discovery of generic community biology that is independent
of actual taxonomic composition. A dataset where these aspects can
be explored is the GlobalPattern dataset of Caporaso \emph{et al.~}\cite{Caporaso2011e}
with 26 samples from human microbiomes, various environments, and
mock communities. We transformed OTU tables to RADs and normalized
them to MaxRank $R=2067$, the minimum richness in the set, and we
computed a distance matrix of all pairs of NRADs.

Hierarchical clustering of the distance matrix (dendrogram in Fig~\ref{fig10}A)
led to close clustering of some samples that also form taxonomic clusters
\cite{Caporaso2011e}, namely the creek samples and a lake sample,
the mock communities, or the human tongue and most feces communities.
The NRADs in these clusters have low distances and thus are similar
(e.g.\ Fig~\ref{fig10}B).

\begin{figure}[!h]
\includegraphics[width=0.8\textwidth]{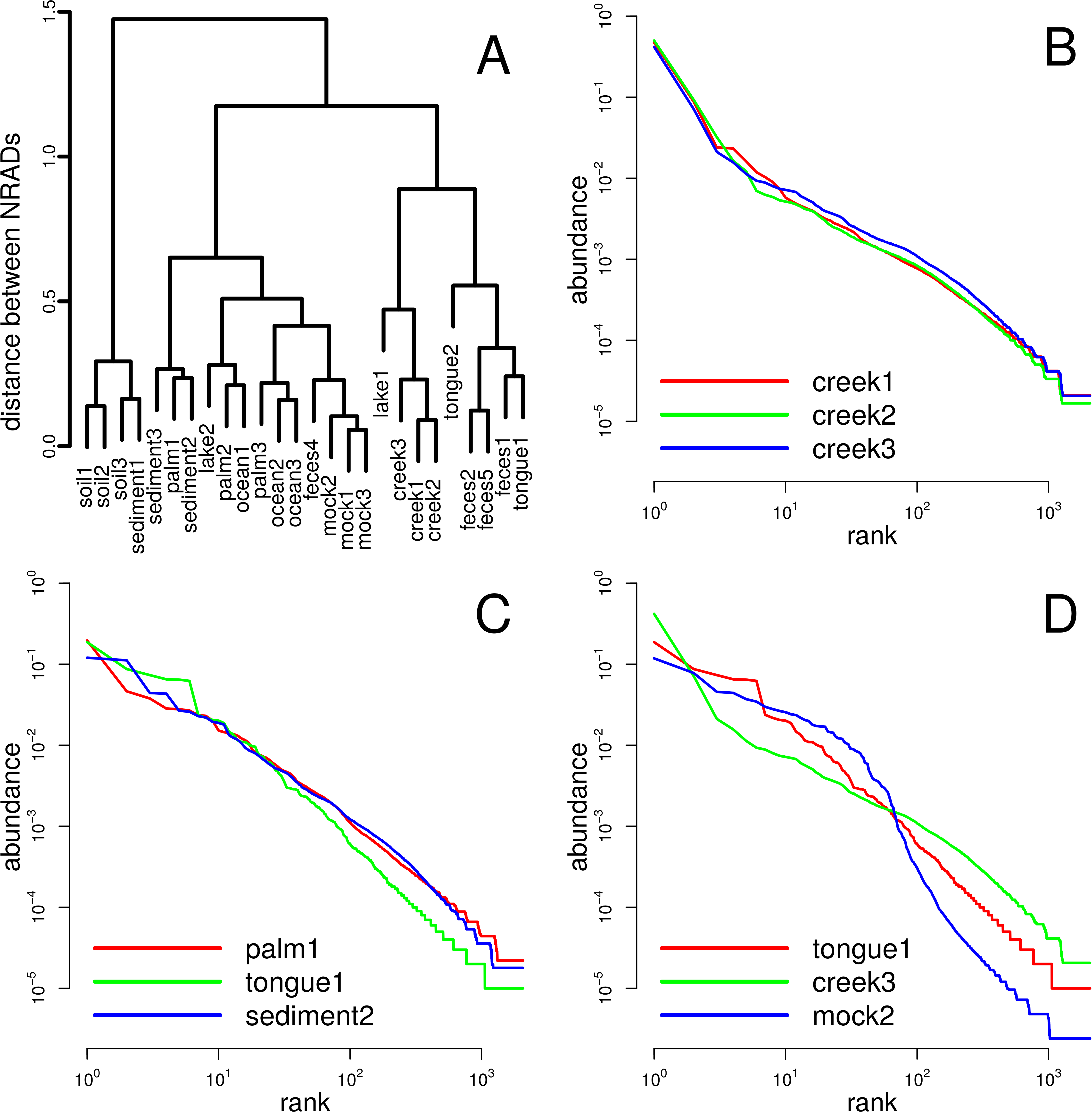}
\caption{\textbf{GlobalPatterns dataset.} (A) Hierarchical clustering dendrogram
based on distances between NRADs. (B) NRADs of three samples of similar
origin that form a cluster. (C) Microbiome of human palm of human
individual 1 clusters closely with sediments 2 and 3 , but is more
distant to tongue microbiome of individual 1. (D) Three differently
shaped NRADs with same entropy.}

\label{fig10} 
\end{figure}

Other relationships are more unexpected. The taxonomy-based analysis
by \cite{Caporaso2011e} clusters tongue and palm microbiomes together
and clearly separates human microbiomes from environmental samples.
Conversely, NRADs of microbiomes of human palms do cluster closer
with environmental samples than with microbiomes of tongues or feces.
For instance, the NRAD \emph{palm1} does not cluster with the microbiome
\emph{tongue1} of the same individual, but most closely with the sediment
samples. NRADs of both \emph{palm1} and \emph{sediment2} have a lower
abundance of rank 1 than tongue1 but more heavy tails (Fig~\ref{fig10}C),
were \emph{palm1} and \emph{sediment2} fit almost perfectly. Reasons
for this unexpected clustering are not known. It could be that microbiomes
in tongue and feces are more strictly controlled by the host body
and the microbiome itself, while microbiomes on palms are more exposed
to the environment.

In the GlobalPatterns dataset, entropies $H_{R}$ range from 2.67
(\emph{tongue2}) to 6.58 (\emph{soil2}) with a median of 4.10 and
a standard deviation of 1.06, and one may be tempted to explain the
observed clustering by different entropies. However, such a simple
approach is not successful here. For instance, the three NRADs in
Fig~\ref{fig10}D have almost the same entropies ($H_{R}^{creek3}=3.98$,
$H_{R}^{mock2}=3.95$, $H_{R}^{tongue1}=3.91$), but have completely
different shapes, and are members of different clusters. In general,
the rich information content of an NRAD cannot be reduced to a single
scalar quantity.

In the introduction we have mentioned that a theoretical possibility
to quantitatively compare pairs of RADs of different richness (a strength
of MaxRank normalization) is the use of the Kolmogorov-Smirnov statistic
$D$, without RAD normalization. This corresponds to using the Chebyshev
distance between the corresponding two cumulative distribution functions
with different support. Although it is obvious that this approach
treats the tail regions in a problematic way, it is unclear whether
this problem is of practical relevance. To test this, we treated $D$
as a distance and reran the hierarchical clustering with this distance.
In fact, this $D$-based tree recovers some of the features of Fig~\ref{fig10}A,
but in general shows less biologically meaningful clusters (\nameref{S1_Fig}).
We conclude that a $D$-based analysis is in practice no alternative
to MaxRank normalization.

As mentioned in the introduction, one commonly used
procedure for quantitative RAD comparisons is a parametric approach.
Typically, a standard distribution such as the log-normal is fitted
to a set of RADs and the comparison is then reduced to a comparison
of the fitted parameters. We have tested the viability of this approach
by fitting to the RADs of all GlobalPatterns samples five standard
distributions, broken stick as null model, preemption, log-normal,
Zipf, and Mandelbrot \cite{Wilson1991}. For most samples, with the
exception of some soil and sediment samples, we found clear \emph{qualitative}
deviations from all five fitted distributions (\nameref{S2_Fig}),
or, in other words, only few fitted distributions reflected the actual
RAD shapes. This means that this parametric approach is in general
not an option for quantitative analysis of HTSeq data. In contrast,
with our non-parametric approach we can quantitatively compare all
information-rich NRADs in a consistent and detailed way.

\subsection*{Robustness of analyses based on MaxRank normalization}

The key feature of MaxRank normalization is that it enables quantitative
RAD comparisons by mapping RADs with diverse richness values to NRADs
of a common richness $R$. It is clear that the lower $R$, the less
information can be carried by NRADs. This could make analyses based
on NRADs sensitive to $R$. To test this, we have tested how the choice
of $R$ affects analyses NRAD based classification, NRAD distances,
and NRAD entropies, as used in the previous sections.

First, we address the question whether NRAD based classification is
sensitive to $R$ using the classification of gut microbiomes described
earlier. We use the same random forest classification and threefold
cross validation as before to classify NRADs of gut microbiomes of
individuals older than 3 years into MV (Malawi/Venezuela) or US (United
States). The results for $R=4105$, $1000$, and $250$ are summarized
in Table~\ref{tab:classification}.

\begin{table}[!ht]
\centering \caption{ \textbf{Accuracy of NRAD-based classification.}}

\begin{tabular}{|l|l|l|}
\hline 
{\boldmath $R$}  & {\boldmath $ACC$}  & {\boldmath $\kappa$}\tabularnewline
\hline 
$4105$  & $0.948\pm0.018$  & $0.862\pm0.049$\tabularnewline
\hline 
$1000$  & $0.947\pm0.022$  & $0.860\pm0.057$\tabularnewline
\hline 
$250$  & $0.917\pm0.025$  & $0.784\pm0.064$\tabularnewline
\hline 
\end{tabular}

\begin{raggedright}
Accuracy $ACC$ and $\kappa$ statistic of NRAD-based classification
for country of origin of gut microbiome as function of MaxRank $R$. 
\par\end{raggedright}

\label{tab:classification} 
\end{table}

The table shows a weak reduction in accuracy and $\kappa$ statistic
with decreasing $R$ and increasing errors, both explainable by a
loss of information stored in the NRADs with lower $R$. However,
the reduction of $R$ from $4105$ to $1000$ affects the accuracy
of the classifier barely, and even the classifier computed from NRADs
with $R=250$ has still a good accuracy. The weak dependency is consistent
with the importance plot (Fig~\ref{fig7}B) where the regions of
high importance cover two extended rank intervals that can be mapped
down to NRADs with $R=250$ or lower, though with some losses.

Secondly, we test the effect of $R$ reduction on NRAD distances for
the Global\-Patterns data set. Fig~\ref{fig11}A demonstrates that
NRAD distances (Eq~\ref{eq:Manhattan}) are well-conserved even if
$R$ is an order of magnitude lower than the richness of the original
RADs. Approximately halving $R$ from $R=2067$ (the maximum possible
$R$ in the GlobalPatterns set) to $R=1000$ does not affect distances
between NRADs: the points are very close to the diagonal (coefficient
of determination $r^{2}=1.000$). If we reduce $R$ more drastically
to $250$ (green points in Fig~\ref{fig11}A), deviations appear,
but we still have $r^{2}=0.976$. The small deviations of NRAD distances
in Fig~\ref{fig11}A for $R=250$ are all towards smaller NRAD distances
(points shifted to the left of the diagonal), because lowering $R$
from $2067$ to $250$ means that for $R=250$ we straighten some
of the fine-grained structure of NRADs with $R=2067$ that allows
for larger NRAD distances. Even for $R=100$ we still have $r^{2}=0.913$.

\begin{figure}[!h]
\includegraphics[width=0.8\textwidth]{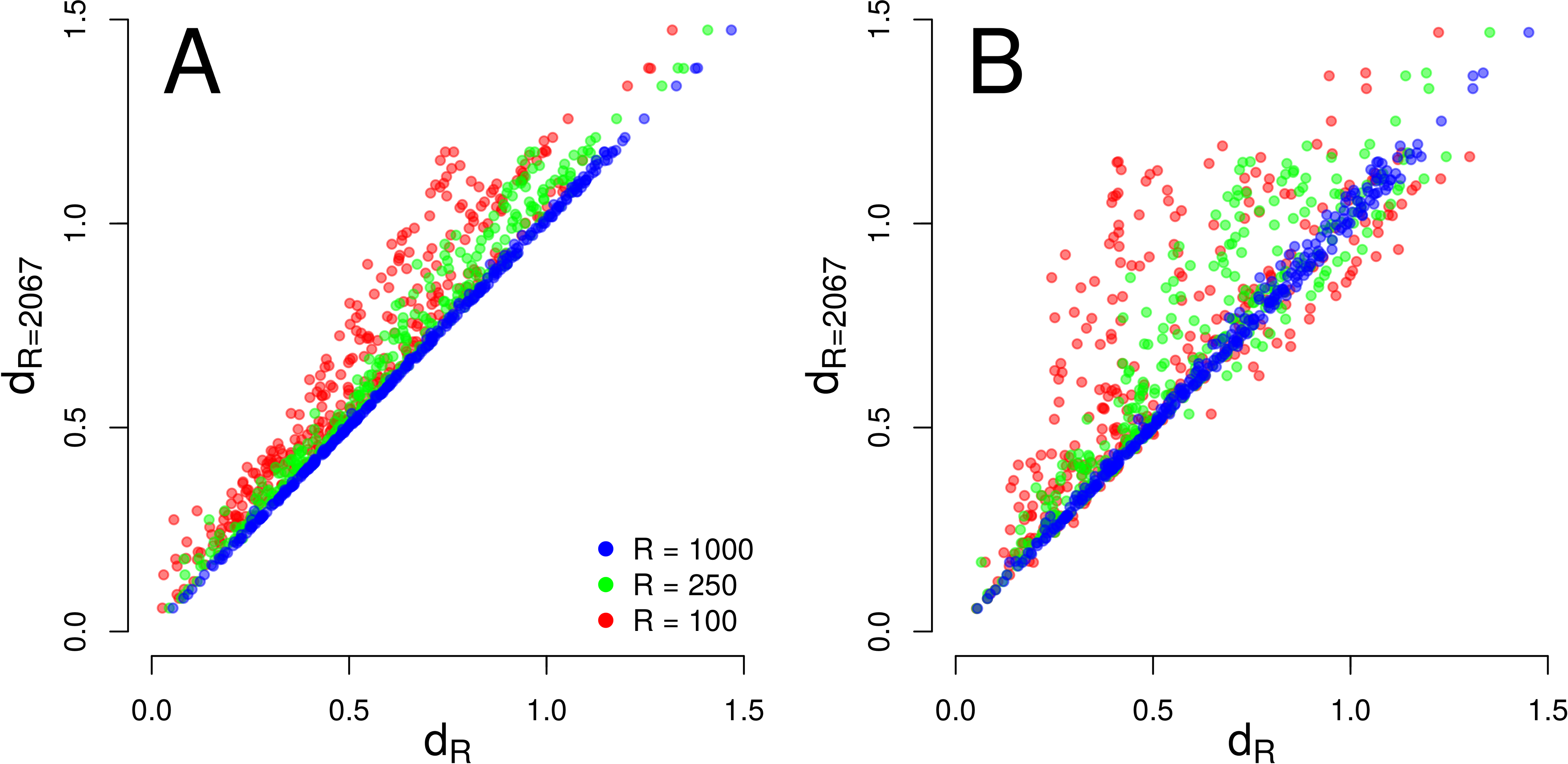}
\caption{\textbf{Dependence of NRAD distances $d_{R}$ on MaxRank $R$.} Co-ordinates
are distances $d_{R}$ between all $26\cdot(26-1)/2=325$ NRAD pairs
of the 26 GlobalPatterns samples at three different values of $R$.
If the distance of an NRAD pair is the same for both $R$, the point
lies on the diagonal. (A) MaxRank normalization; (B) cutoff normalization.}

\label{fig11} 
\end{figure}

In Fig~\ref{fig11}B we show for comparison how reduction of $R$
affects NRAD distances for a simpler normalization scheme in which
ranks above a given $R$ are cut off (``cutoff normalization''),
i.e.\ information in the tails with ranks higher than $R$ is completely
neglected. The scatter of the points is generally wider for all three
values of $R$ with $r^{2}=0.995$, $0.874$ and $0.614$, respectively.
These lower $r^{2}$ values point to the importance of the tails that
are neglected by cutoff normalization. NRAD pairs with differences
predominantly in the neglected tails appear in Fig~\ref{fig11}B
shifted to the left of the diagonal, while NRAD pairs with more divergent
heads and more similar tails appear right shifted. Cutoff normalization
clearly is less robust than MaxRank normalization with respect to
the choice of $R$.

As a consequence of the robustness of NRAD distances, the down-stream
analyses that make use of NRAD distances are also quite robust. For
instance, the hierarchical clustering (Fig~\ref{fig10}A) of the
GlobalPatterns dataset is identical between $R=250$, $R=1000$, and
$R=2067$ up to an agglomeration height of 0.98 (vertical axis in
Fig~\ref{fig10}A). Cluster assignments differ only above this agglomeration
height at the highest branching points and thus at the most fuzzy
cluster level.

Thirdly, we find for each sample $i$ a regular and systematic $R$
dependence of NRAD entropy $H_{R}^{(i)}$ (Eq~\ref{eq:Shannon}):
\begin{equation}
H_{R}^{(i)}/\log R\approx\textrm{c}^{(i)}.\label{eq:HofR}
\end{equation}

with a sample dependent constant $c^{(i)}$. This approximation works
reasonably well if $R$ is not varied by more than an order of magnitude.
Eq~(\ref{eq:HofR}) corresponds formally to the definition of Shannon
evenness with $R$ interpreted as richness (Eq~\ref{eq:Shannon}),
i.e.\ Shannon evenness changes weakly with $R$. Moreover, this equation
means that entropy or information content will systematically decrease
with decreasing $R$. This decrease will depend weakly on $R$ since
$\log R$ changes only slowly with $R$.

As the NRAD entropies change systematically with $R$, this must also
affect our model of entropy of gut microbiome NRADs as function of
age. For instance if we approximately halve $R$ from $4105$ to $2000$
and repeat the fitting of the model Eq~(\ref{eq:H_of_t}) we arrive
at $H_{2000}^{0,MV}=3.34\pm0.16$, $H_{2000}^{0,US}=3.35\pm0.19$,
$\lambda_{2000}^{MV}=0.70\pm0.09\, yr^{-1}$, $\lambda_{2000}^{US}=1.18\pm0.27\, yr^{-1}$,
$H_{2000}^{max,MV}=5.62\pm0.06$, and $H_{2000}^{max,US}=5.12\pm0.03$.
Entropic model parameters $H_{2000}$ are systematically lower by
a small amount than the corresponding $H_{4105}$ values (Table~\ref{tab:entropy-age-parms})
as expected from an approximate scaling (Eq~\ref{eq:HofR}). If all
entropic factors in Eq~(\ref{eq:H_of_t}) scale in the same way,
the scaling factor cancels, and the exponential growth term $e^{-\lambda t}$
should stay the same. In fact, the growth parameters $\lambda$ are
almost unaffected by the decrease of $R$ from $4105$ to $2000$.

Finally, one important aspect of robustness is the following. If we
take several samples of different size and therefore different richness
of the same, well-mixed generalized community, then
determine the RADs of those samples, and finally normalize these RADs
to the same $R$, we should obtain the same NRAD for all samples.
Only if this requirement is met can NRADs inform reliably about a
generalized community. To test fulfillment of this
requirement, we first down-sampled original HTSeq data sets by an
order of magnitude in richness. Fig~\ref{fig12}A shows as an example
the original RAD and the down-sampled RAD of the first sample in the
gut microbiome data of \cite{Yatsunenko2012} (MG-RAST ID 4.4899263e6).
We then normalized both to the same MaxRank $R=1000$ (Fig~\ref{fig12}B).
For all samples we found that both RADs, the original and the down-sampled,
led to practically the same NRAD. The violin plot Fig~\ref{fig12}C
makes a quantitative statement about this property: the biologically
relevant distance distributions between NRADs are the same for both
the original and down-sampled data (left and middle violin in Fig~\ref{fig12}C
are equal). In comparison, the distances between corresponding pairs
of NRADs of original and down-sampled RADs are negligible (right violin
in Fig~\ref{fig12}C) as it should be for a normalization procedure
that is robust against the size of the source sample.

\begin{figure}[!h]
\includegraphics[width=0.8\textwidth]{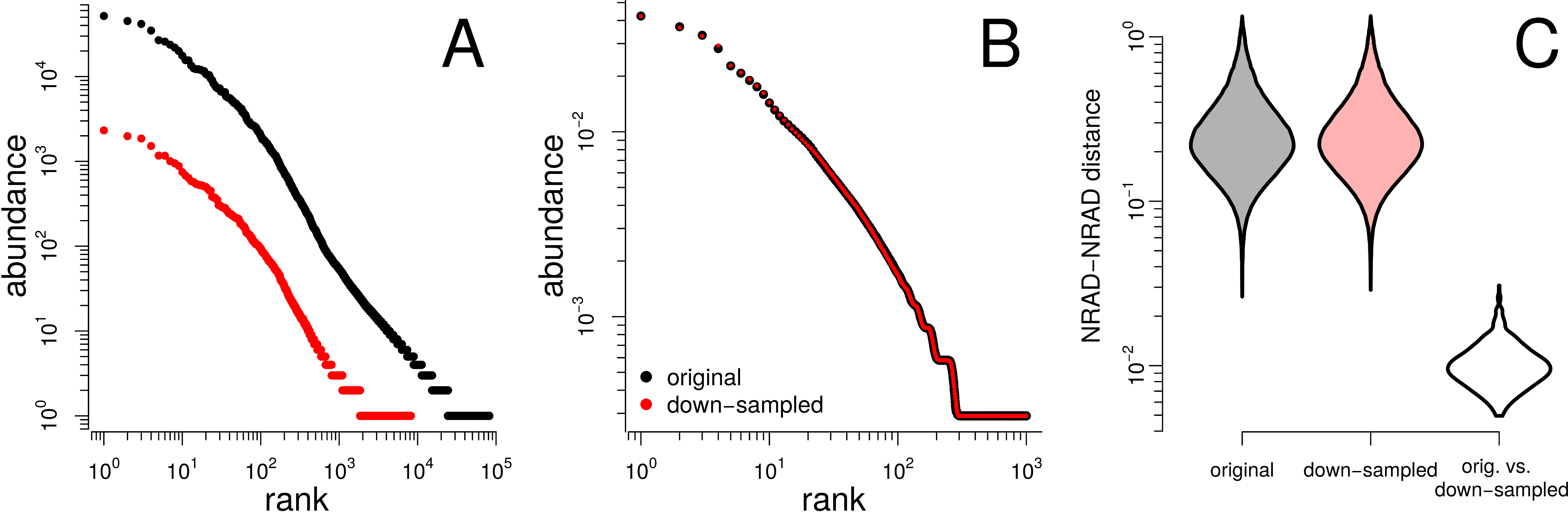}
\caption{\textbf{Robustness of NRADs against varying sampling depth.} (A) original
RAD of first sample of \cite{Yatsunenko2012} (black) and down-sampled
RAD (red). (B) the two NRADs obtained by MaxRank normalization to
$R=1000$ of the RADs in panel A are almost indistinguishable. (C)
comparison of NRAD distances of the first 50 samples of the data set
of \cite{Yatsunenko2012}. Left violin plot: density of distances
between NRADs computed by MaxRank normalization to $R=1000$ of the
original RADs; middle violin plot: same for down-sampled RADs; right
violin plot: distances between corresponding original and down-sampled
NRADs. The biologically meaningful NRAD distance distributions are
robust against differences in sample size (left and middle violin).
In comparison, the distances related to differences in sample size
are negligible (right violin).}

\label{fig12} 
\end{figure}

In summary, we found that NRADs are robust quantitative descriptors
of RADs. However, it is also clear that there are critical values
of $R$ below which essential RAD structures are lost and analyses
become inconclusive. These critical values will depend on the studied
RADs and on the analysis method. We recommend to monitor NRAD quality
with methods suitable for the respective question. For instance, if
the cluster structure a set of NRADs is of interest, clustering and
ordination methods as those presented above can be used to detect
loss of structure when results for several $R$ values are compared.

\section*{Discussion}

MaxRank normalization makes communities of different richness quantitatively
comparable by mapping their RADs to NRADs of a common richness $R$.
This is similar to projecting objects of different higher dimensions
to one common lower dimension where the projections can be compared
directly. The price to be paid is information loss,
especially loss of richness information. However, the remaining information
enables new approaches to community analysis. In particular, we could
show that the information extracted from NRADs is sufficient to generate
quantitative models of the dynamics or composition of generalized
communities.

MaxRank normalization is not the only possible algorithm that maps
higher richness RADs to a common lower richness, but it has crucial
advantages over other procedures. For instance, another procedure
is to cut off in all RADs ranks beyond a common maximum rank $R$.
Obviously, this simple procedure neglects information in the RAD tails,
leading to lower robustness as shown earlier. An alternative
would be to scale the rank axis to the same maximum number. This would
lead to fractional pseudo-ranks; the corresponding pseudo-rank abundance
vectors have different dimensions and thus cannot be compared directly.
Another alternative is the coarse-graining of the rank axis to a given
number of pseudo-ranks. This is also not satisfying as the coarse-graining
would treat samples of different richness differently, and because
the pseudo-ranks are not observables. Conversely, MaxRank normalization
is conceptually attractive since it corresponds to a real experimental
sampling process with the attainment of a given richness as stop criterion.
In fact, a similar observational protocol (``m-species
list''), i.e.~sampling up to a constant maximum rank, has been used
in field ecology \cite{Mackinnon1993}. MaxRank normalization does
not impose a specific model, i.e.\ the approach is model-free and
generally applicable. For instance, we have applied it not only to
HTSeq data but also to data from conventional ecological sampling
of macro-invertebrates from fresh-water systems (unpublished).

MaxRank normalization (re-sampling up to a given maximum
rank) may be confused with \emph{rarefying} (re-sampling up to a given maximum number of individuals) \cite{Koren2013,McMurdie:2014}.
However, the two methods answer different diversity questions: rarefying
allows answering questions that relate to sample richness (and typically
also to OTU abundance), while NRAD comparison largely eliminates richness
(and OTUs) and puts emphasis on abundance structure difference.

There are a number of limitations of MaxRank normalization. First,
it is obvious that most information about richness is lost. Interestingly,
it is not completely lost, but partially encoded in the NRADs, as
a detailed technical analysis shows (to be published elsewhere). Nevertheless,
if mainly changes or differences in richness are of interest, NRADs
are not suitable.

Second, and related to the previous limitation, it is possible that
some of the samples to be compared are richness-limited while others
are not. In this case the information loss mentioned above can turn
the method useless. To illustrate this point, imagine two systems that
should be compared, one with two species, the other with thousands
of species. In this case we would normalize the system to $R=2$ and
lose almost all the information in the RAD of the richer system. This
limitation is usually not serious for the analysis of HTSeq data of
generalized communities such as microbiomes. One
strategy to cope with outlier samples of extremely low richness that
would enforce the use of a low $R$ and severe loss of information
in NRADs is to discard such outliers, thus sacrificing some breadth
for higher accuracy.

Third, a more severe problem with HTSeq data is the often implied
assumption that read counts are quantitative measures of abundances.
Unfortunately, HTSeq data can be biased by the experimental protocol,
e.g.\ by preferential PCR amplification of certain species and non-amplification
of others, or it can contain false positives, e.g.\ error mutants
or chimeras produced in the experimental process \cite{Schloss2011,Degnan2012}.
Since RADs are derived from OTU tables, any abundance
bias that affects OTU tables will also affect RADs. Although these
problems do not limit applicability of MaxRank normalization, they
do limit the possible biological interpretation of the results. Related
to this point, we found that changes in HTSeq protocols can significantly
impact RADs and therefore NRADs. Thus, a stringent control is required
if results from different studies are to be compared. Such a control
can be implemented e.g.\ by comparing RADs resulting from the application
of the different protocols to common reference samples. As HTSeq technology
develops rapidly, some of these experimental problems may be solved
in the near future. But even with these imperfections, NRADs can be
used as generic quantitative descriptors to discover new community
biology.

Fourth, analyses based on NRADs alone are blind to taxonomic composition. This can be an advantage because in this way generic effects that influence the abundance structure can become clearly visible. But this blindness to taxonomy makes NRAD based analyses inadequate if differences or changes in taxonomic composition are of major interest.

\section*{Supporting Information}


\paragraph*{S1 Fig.}

\label{S1_Fig} \textbf{Kolmogorov-Smirnov statistic $D$ as RAD-RAD
distance.} Hierarchical clustering dendrogram of GlobalPatterns RADs,
using Kolmogorov-Smirnov statistic $D$ between pairs of non-normalized
RADs as distance {(results with the Anderson-Darling
statistic were essentially the same, not shown)}. Some of the samples
(e.g.\ \emph{soil1} and \emph{soil2}) are clustered similar to the
NRAD-NRAD distance based dendrogram Fig~\ref{fig10}A of main text.
More often, clustering based on NRAD distances is biologically more
meaningful than $D$-based clustering (e.g.\ \emph{ocean1/2/3}).
In some cases, the $D$-based clustering is biologically outright
wrong as for the cluster formed by the high-evenness mock samples
and the low-evenness tongue samples.

\paragraph*{S2 Fig.}

\label{S2_Fig} {\textbf{Fits of four standard distributions to RADs
of GlobalPatterns set.}}

\paragraph*{S1 Table.}

\label{S1_Table} \textbf{Meta-data for GlobalPatterns and correspondence
of original sample IDs and new sample names.}


%


\end{document}